\documentclass{article}

\PassOptionsToPackage{numbers, compress}{natbib}
\usepackage[main, final]{neurips_2026}
\usepackage[utf8]{inputenc}
\usepackage[T1]{fontenc} 
\usepackage{hyperref}      
\usepackage{url}       
\usepackage{booktabs}     
\usepackage{amsfonts}     
\usepackage{nicefrac}     
\usepackage{microtype}    
\usepackage{xcolor}         
\usepackage{amsmath}
\usepackage{algorithmic}
\usepackage{array}
\usepackage{subfig}
\usepackage{textcomp}
\usepackage{stfloats}
\usepackage{verbatim}
\usepackage{graphicx}
\usepackage{amssymb}
\usepackage{xspace}
\usepackage{multirow}
\usepackage{tabularx}
\usepackage{tablefootnote}
\usepackage{makecell}
\usepackage{pifont}
\usepackage{enumitem}
\usepackage{colortbl}
\usepackage{placeins}
\usepackage{wrapfig}
\usepackage{subcaption}

\definecolor{mygreen}{RGB}{0, 150, 0}
\definecolor{myred}{RGB}{200, 0, 0}
\definecolor{lightgray}{gray}{0.92}

\newcommand{\cmark}{\textcolor{mygreen}{\ding{51}}}
\newcommand{\xmark}{\textcolor{myred}{\ding{55}}}

\newcommand{\tool}{\textsc{AnyEdit}\xspace}
\newcommand{\toolb}{\textsc{AnyEditBench}\xspace}

\title{\tool: A Unified Framework for Speech and Singing Voice Editing with Real-World Environmental Consistency}

\author{%
  Yunjia Zhang$^{1}$\thanks{Equal contribution.},
  Junan Zhang$^{1}$\footnotemark[1],
  Jing Yang$^{2}$\footnotemark[1],
  Xueyao Zhang$^{1}$, Fan Fan$^{2}$,
  Zhizheng Wu$^{1,3,4}$\thanks{Corresponding author.
  E-mail: \href{mailto:yunjiazhang2@link.cuhk.edu.cn}{yunjiazhang2@link.cuhk.edu.cn},
  \href{mailto:wuzhizheng@cuhk.edu.cn}{wuzhizheng@cuhk.edu.cn}} \\
  $^1$The Chinese University of Hong Kong, Shenzhen \\
  $^2$Institute of Media Technology and Experience Design, Huawei \\
  $^3$City University of Macau \\
  $^4$Amphion Technology Co., Ltd. \\
}

\begin{document}

\maketitle

\begin{abstract}
We present \tool, a unified, separation-free framework for speech and singing voice editing that preserves real-world acoustic environments, including noise, reverberation, and background music. Recent unified generative models already achieve strong speech and singing generation in clean conditions, but real-world vocal editing remains challenging because the model must revise content while simultaneously preserving speaker identity, prosody, and the surrounding acoustic scene. A straightforward solution is to directly model noisy edited audio end to end, yet this entangles semantic editing with environmental acoustics and can weaken the core vocal generation capability. We therefore adopt a two-stage design that preserves clean-domain editing ability while deferring acoustic rendering to a separate stage: an instruction-guided autoregressive editor first predicts clean prosody and content-style tokens, supported by a teacher-student noise-robust prosody tokenizer that distills clean prosodic tokens from degraded recordings; a flow-matching acoustic model then performs in-context rendering conditioned on the source acoustic context, so the edited segment inherits surrounding noise, reverberation, and background sound while maintaining consistent speaker timbre. To stabilize unedited regions, we introduce unmodified-area-aware supervision; after pre-training, we further apply direct preference optimization (DPO) on mined real-to-real singing pairs via Chromagram-DTW. We also release \toolb, a benchmark covering both speech and singing voice editing under clean, simulated, and in-the-wild acoustic conditions. Experiments on Chinese and English speech and singing datasets showed that \tool consistently outperforms strong baselines in content accuracy, subjective quality, and background consistency under noisy and reverberant conditions. Demo audios can be found at \url{https://any-edit.github.io/}. \toolb is released at \url{https://huggingface.co/datasets/Ririiiii/anyeditbench}.
\end{abstract}

\section{Introduction}
Audio synthesis technologies, such as text-to-speech (TTS) and text-to-singing (TTSing), have achieved remarkable progress in generating high-quality vocals from scratch. Recent unified generative models, such as Vevo2~\cite{zhang2025vevo2}, can simultaneously handle speech and singing generation with impressive fidelity under clean conditions. However, in practical audio post-production for user-generated content (UGC), podcasts, and music, creators frequently need to perform zero-shot localized editing on an existing recording—such as inserting, deleting, or substituting words and lyrics (e.g., correcting mispronunciations or updating lyrics)—rather than synthesizing a new one entirely. In real-world scenarios, the original audio is rarely clean; vocals are inevitably entangled with complex background acoustics, including noise, reverberation, and background music.

Existing approaches to real-world vocal editing fall into two paradigms, neither of which fully addresses this challenge. \textbf{The first paradigm is direct end-to-end editing}, which couples semantic editing and acoustic-scene reproduction within a single generative stage. Speech editors such as VoiceCraft~\cite{peng2024voicecraft} and Step-Audio-EditX~\cite{yan2025step}, singing editors such as EditSinger~\cite{zhang2022editsinger} and SongEditor~\cite{yang2025songeditor}, and general-purpose unified audio models such as Ming-UniAudio~\cite{yan2025ming} and MiMo-Audio~\cite{zhang2025mimo} all follow this approach. Since paired (noisy source, noisy target) supervision is essentially unobtainable at scale, these methods inherit a clean-domain prior and lose intelligibility, speaker fidelity, or background continuity on real-world inputs; the unified models additionally lack localized editing support. \textbf{The second paradigm is the ``separation-edit-mix'' pipeline}, which isolates the clean vocal via source separation, edits it with TTS/TTSing, and remixes it with the background. This pipeline introduces three intrinsic issues, one per stage: (1) \textbf{Separation artifacts}, especially severe for singing where the background is harmonically correlated music rather than uncorrelated noise~\cite{jiang2025singverse}; (2) \textbf{Static background modeling}, as methods such as SeamlessEdit~\cite{chen2025seamlessedit} (which directly sums the edited vocal with the separated background) and DAIEN-TTS~\cite{lu2025daien} (which conditions synthesis on a single global SNR) cannot follow time-varying background dynamics such as fade-ins, fade-outs, or acoustic gaps; and (3) \textbf{Lack of true localized editing}, since the edit step is typically realized by TTS-style models that regenerate the entire utterance rather than modifying only the target span---SSR-Speech~\cite{wang2025ssr}, for instance, regenerates unedited regions instead of keeping them, failing to maintain the fine-grained acoustic context of the original recording. As a result, unified speech and singing editing under realistic acoustics remains insufficiently explored; we provide a comprehensive survey in Appendix~\ref{app:related}.

This raises a central question: \textbf{\textit{can a single framework achieve speech and singing voice editing that is unified across both domains, noise-robust on the input side, and environment-preserving on the output side?}}

We propose \tool, a separation-free two-stage framework that addresses this question. \textbf{Our key insight is that semantic editing (what to say) and acoustic rendering (how it sounds in context) are fundamentally separable}: semantic editing benefits from clean-domain training, while acoustic rendering only requires faithful in-context reproduction. This decoupling lets \tool inherit the editing precision of clean-trained unified models while avoiding source separation entirely. The framework realizes this insight through three coordinated stages. \textbf{(1) Noise-Robust Prosody Tokenizer.} A teacher-student tokenizer distills clean prosody tokens from noisy chromagrams, providing stable prosodic conditions for the downstream editor. \textbf{(2) Instruction-Guided Autoregressive (AR) Editor.} The AR editor predicts clean prosody and content-style tokens from either target text or natural-language instructions, with \textit{unmodified-area-aware supervision} that upweights the loss on prosody tokens within unchanged spans to prevent prosody drift. We further refine singing quality via Direct Preference Optimization (DPO) on mined \textit{real-to-real} pairs from music structural repetitions (e.g., different verses sharing the same prosody but different lyrics), avoiding the synthesis artifacts that SVS-generated targets would otherwise inject into the editor. \textbf{(3) Flow-Matching Acoustic Model (FMT).} A flow-matching transformer renders the target mel-spectrogram conditioned on the source acoustic context, so the edited segment inherits surrounding noise, reverberation, and background music while preserving speaker timbre.

To support realistic evaluation, we introduce \toolb, a benchmark for real-world vocal editing comprising 300 speech and 300 singing entries across diverse acoustic scenarios (clean, simulated, and in-the-wild recordings). Extensive subjective and objective experiments on Chinese and English datasets demonstrate that \tool consistently outperforms strong baselines in content accuracy, subjective quality, and background continuity. Our main contributions are summarized as follows:

\vspace{-2mm}
\begin{itemize} [itemsep=0ex,leftmargin=3ex]
\item We propose \tool, the first unified framework for speech and singing voice editing in real-world acoustic environments. By decoupling semantic editing from acoustic rendering, \tool combines a noise-robust semantic stage with an environment-aware in-context rendering stage, eliminating the need for source separation.
\item We introduce a suite of separation-free techniques: (i) a noise-robust prosody tokenizer via teacher-student distillation; (ii) an instruction-guided AR editor with \textit{unmodified-area-aware supervision}; (iii) DPO alignment on \textit{mined real-to-real singing pairs} via Chromagram-DTW matching of repeated song structures; and (iv) an in-context flow-matching acoustic renderer.
\item We establish \toolb, the first benchmark for real-world vocal editing in Chinese and English across both speech and singing under clean, simulated, and in-the-wild conditions, on which \tool achieves state-of-the-art performance against existing baselines under noisy and reverberant settings in content accuracy, speaker fidelity, and background consistency
\vspace{-2mm}
\end{itemize}

\section{Method}
\begin{figure}[t]
  \centering
  \includegraphics[width=1\linewidth]{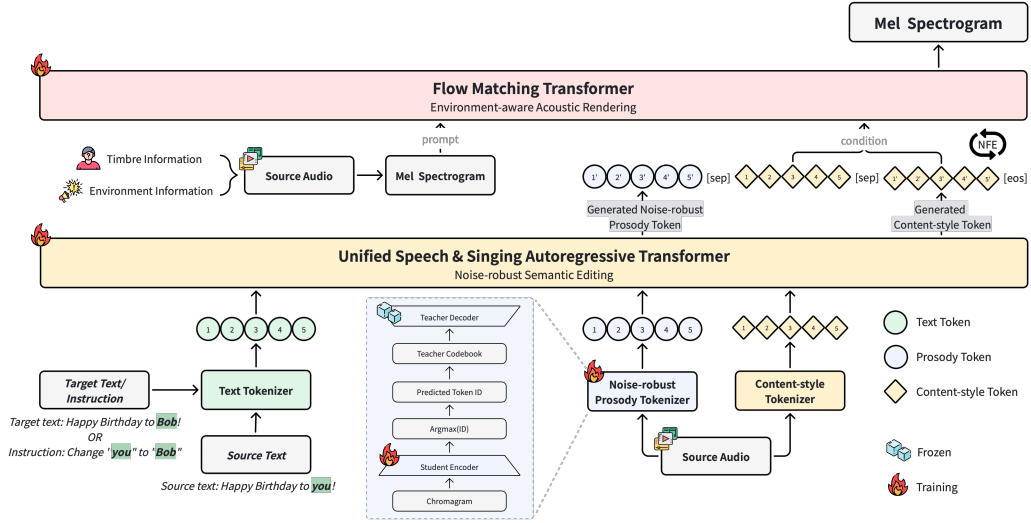}
  \caption{Overall architecture of the \tool framework. NFE indicates the Number of Function Evaluations (ODE integration steps) used during flow matching inference.}
  \label{fig:framework}
\vspace{-4mm}
\end{figure}

We formulated real-world speech and singing editing as a conditional generation problem: given a source recording, its transcript, and either a target transcript or an instruction, the system synthesized target vocals that precisely incorporated the semantic edits while preserving the unedited vocal context and original acoustic environment. As shown in Figure~\ref{fig:framework}, \tool achieved this via a separation-free, cascaded architecture: (1) an AR model first generated discrete prosody tokens and content-style tokens in a clean latent space, (2) followed by a FMT that rendered the target mel-spectrogram conditioned on the original timbre and acoustic context.
\vspace{-4mm}

\subsection{Noise-Robust Prosody Tokenizer}
\begin{wrapfigure}{r}{0.5\linewidth}
  \centering
  \vspace{-10pt}
\includegraphics[width=\linewidth]{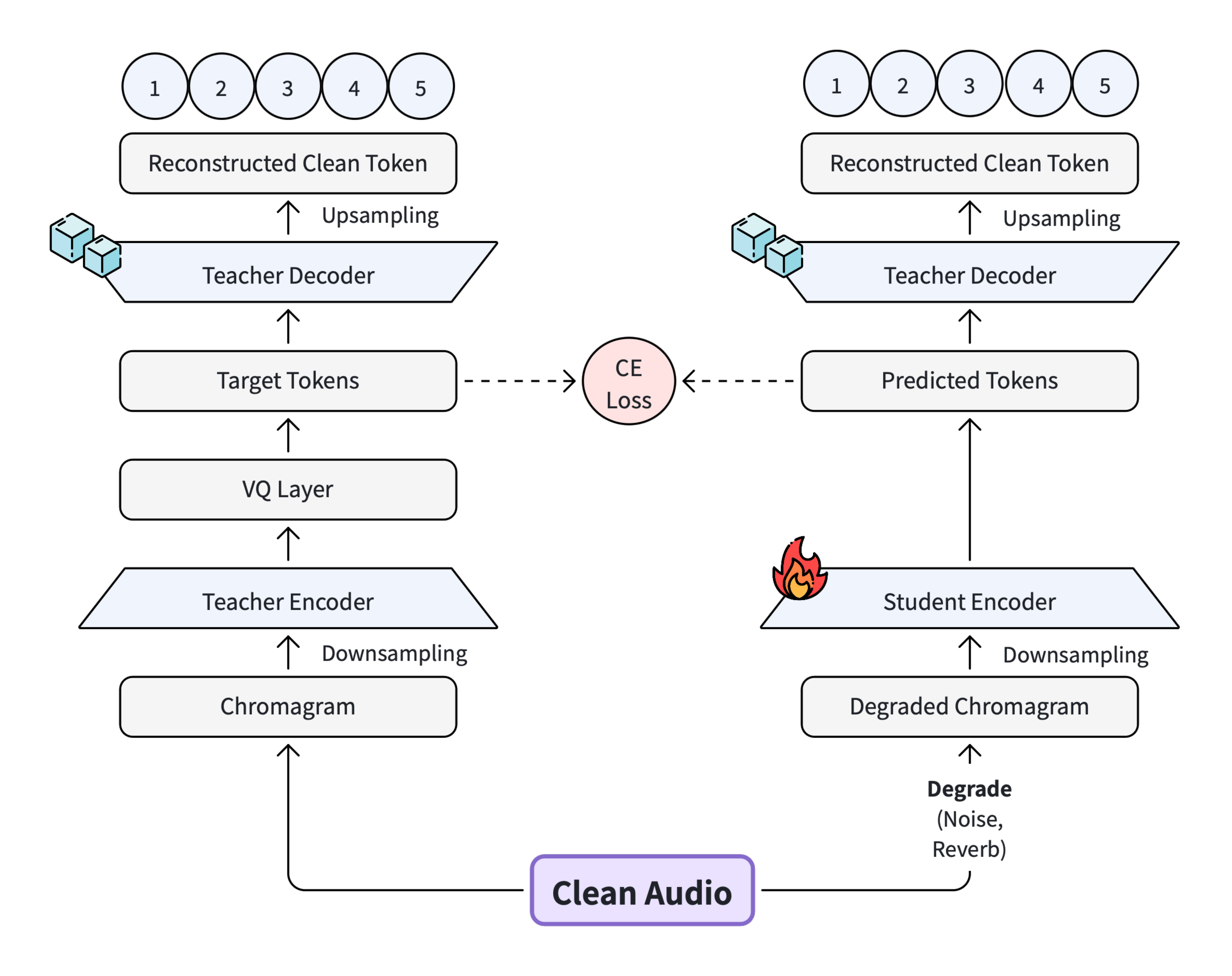}
  \caption{Training pipeline of the Noise-Robust Prosody Tokenizer via Teacher-Student Distillation.}
  \label{fig:tokenizer}
  \vspace{-10pt}
\end{wrapfigure}

Real-world recordings often contain noise, reverberation, and background music, making prosody extraction unreliable. To provide stable prosodic conditions for both speech and singing editing, we built a noise-robust prosody tokenizer. Rather than relying on explicit source separation, which typically introduces artifacts and disrupts the acoustic correlation between vocals and background, our model directly extracts the prosody of the clean vocal content from the input recordings.

\textbf{Representation Backbone.} We took chromagrams as the tokenizer input. Chromagrams preserve melodic and rhythmic structures while being relatively insensitive to timbre variation and background interference, making them highly suitable for the unified modeling of speech and singing prosody under degraded conditions.

\textbf{Teacher-Student Architecture.} We trained the tokenizer using teacher-student distillation (Figure~\ref{fig:tokenizer}). The teacher tokenizer $Q_{\text{tea}}$ was a VQ-VAE trained on clean audio. Given clean chromagram features $\mathbf{c}_{\text{clean}}$, it produced discrete prosody tokens $z^{\text{tea}}$. The student model $f_{\phi}$ shared the same architecture but took degraded chromagrams $\mathbf{c}_{\text{degraded}}$ as input and was trained to recover the teacher's clean tokens.

The teacher followed a standard VQ-VAE design~\cite{zhang2025vevo2}: it included a downsampling layer, a convolutional encoder, a vector quantization (VQ) module with codebook $\mathbf{E} =[\mathbf{e}_1, \mathbf{e}_2, \dots, \mathbf{e}_K]$, a decoder, and an upsampling layer. The VQ operation mapped encoded features $\mathbf{z}_e(\mathbf{x})$ to the nearest codebook vector $\mathbf{z}_q(\mathbf{x}) = \arg \min_j \|\mathbf{z}_e(\mathbf{x}) - \mathbf{e}_j\|_2$. We extracted 50~Hz chromagram features from raw audio and applied 8$\times$ downsampling, yielding a 6.25~Hz discrete sequence.

To ensure robustness, we trained the student model $f_{\phi}$ via a cross-entropy objective $\mathcal{L}_{\text{tokenizer}} = \sum_{t=1}^{T} \mathrm{CE}(\hat{z}_t, z^{\text{tea}}_t)$, where $z^{\text{tea}} = Q_{\text{tea}}(\mathbf{c}_{\text{clean}})$ and $\hat{z} = f_{\phi}(\mathbf{c}_{\text{degraded}})$. This distillation objective forced the student to recover clean prosody tokens from degraded inputs. While the teacher tokenizer applied to degraded inputs achieved a token match rate of only 0.6081, our student improved this to 0.7037, demonstrating improved noise-robust prosody recovery. The degradation process followed the pipeline of AnyEnhance~\cite{zhang2025anyenhance}.

\subsection{Autoregressive Editing Model (AR)}
The AR model served as the clean semantic stage of \tool. It predicted two streams of discrete tokens: prosody tokens for encoding rhythm and temporal variations, and content-style tokens for encoding linguistic content and vocal styles.

For the content-style tokenizer, we adopt the RepCodec structure~\cite{zhang2025vevo2}, which quantizes continuous self-supervised speech representations into discrete tokens that capture fine-grained phonetic and stylistic nuances while stripping out speaker timbre.

\textbf{Training Strategy.} The AR model was trained in three stages: Phase~I pre-training, Phase~II pre-training, and Direct Preference Optimization (DPO). Since directly learning to predict prosody without time alignment from scratch prevented convergence, we adopted a two-phase curriculum before DPO:
\vspace{-2mm}
\begin{itemize}
    \item \textbf{Phase~I (Content-Style only):} The model predicted only content-style tokens, learning a stable time alignment with the input audio and the mapping from text to vocal content and style.
    \item \textbf{Phase~II (Content-Style + Prosody):} Building upon the alignment from Phase~I, the model jointly predicted content-style and prosody tokens, enabling successful prosody prediction without being fully bound to the original time alignment.
\end{itemize}
\vspace{-2mm}
\begin{figure}[h]
  \centering
\includegraphics[width=1\linewidth]{./fig/ar-training-table.pdf}
  \caption{AR model pretraining data construction and training process}
  \label{fig:ar_stage2}
\end{figure}

\textbf{Pre-training Data Construction.} To scale training without manual annotation, we constructed paired editing data using CTC forced aligner \footnote{\url{https://github.com/MahmoudAshraf97/ctc-forced-aligner}}. We utilized the Emilia dataset~\cite{he2024emilia} for speech and SingNet~\cite{gu2025singnet} for singing. For each source utterance, we retrieved a reference sample from the same speaker following the pipeline of MimicLM~\cite{feng2026mimiclm}. We then synthesized insertion, deletion, and substitution edits at the beginning, middle, or end of the utterance. In total, we processed approximately 1.2 million same-speaker speech pairs for Chinese and English, yielding about 3,000 hours per language, and roughly 2,000 hours of singing paired data per language. The training process is showed in Figure~\ref{fig:ar_stage2}.

\textbf{Model Architecture and Conditioning} 
During training, the target text was used in 50\% of cases, with a natural-language instruction used in the remaining 50\%, allowing the model to perform editing flexibly from either explicit target text or free-form instructions. Detailed Implementations are in Appendix~\ref{appendix:imp_detail}.

\subsubsection{Unmodified-Area-Aware Supervision}

A common error in AR-based vocal editing is prosody drift, where unedited words around the edited span change in timing or intonation. To reduce this effect, we introduced \textit{unmodified-area-aware supervision}.

During training, we computed the Longest Common Subsequence (LCS) between the source and target texts to identify unchanged spans. For contiguous unchanged spans longer than $L_{\min}=4$, we mapped the corresponding text spans to their temporal positions in the source prosody-token sequence. Then we upweighted the negative log-likelihood of tokens in these mapped regions with a penalty factor $\lambda_{\text{keep}}=5.0$:
\begin{equation}
\mathcal{L}_{\text{AR}} = - \sum_{t=1}^{T} w_t \log P_\theta(y_t | y_{<t}, \mathbf{X}),
\end{equation}
where $w_t = \lambda_{\text{keep}}$ if step $t$ belonged to a kept span, and $w_t = 1.0$ otherwise. This encouraged the model to preserve the prosody of unedited content and significantly improved boundary stability during singing editing.

\subsubsection{Preference Alignment via mined Real-to-Real pairs}
\label{app:dpo_data}

A key difficulty in singing voice editing is obtaining paired data in which the source and target share timbre, rhythm, but differ in lyrics. Existing approaches construct such pairs using singing voice synthesis (SVS) models, but synthetic targets inevitably introduce artifacts, biasing the editing model toward synthetic acoustic patterns. See Table~\ref{tab:singing_data_construction} for a comparison with data construction pipelines for singing voice editing. 

\vspace{-2mm}
\begin{table}[h]
\centering
\caption{Comparison of data construction pipelines for singing voice editing. \textbf{Target Realness}: Whether the ground-truth target audio is a natural human recording. \textbf{Melodic Match}: Whether the target strictly follows the source melody contour.}
\label{tab:singing_data_construction}
\renewcommand{\arraystretch}{1.5}
\resizebox{\textwidth}{!}{%
\begin{tabular}{l p{5cm} c c c p{4.5cm}}
\toprule
\textbf{Method} & \textbf{Construction Protocol} & \textbf{Target Realness} & \textbf{Melodic Match} & \textbf{Scalability} & \textbf{Primary Limitation} \\
\midrule
\textbf{Web Crawled Hand-crafted SVS} &
Content creators manually tune synthesis software parameters to align specific timbres with different lyrics. &
\xmark & \cmark & \xmark &
Extremely labor-intensive; dependent on human tuning skills; limited data volume. \\

\textbf{SVS Model Generation} &
Synthesizes target audio from lyrics using a trained SVS model (Synthetic Target). &
\xmark & \cmark & \cmark &
Model learns \textit{synthetic artifacts} and robotic prosody; upper bound limited by the SVS teacher. \\

\textbf{Random Truncation} &
Randomly masks/cuts audio segments for reconstruction (pre-training style). &
\cmark & \xmark & \cmark &
Lacks specific lyrical control; cannot guarantee melodic preservation when lyrics change. \\

\midrule
\rowcolor{lightgray}
\textbf{Mined Real-to-Real Pairing (Ours)} &
\textbf{Mines intra-song repetitions (Segment A/B) via Chromagram DTW alignment.} &
\cmark & \cmark & \cmark &
Requires songs with repetitive structures (common in pop music). \\
\bottomrule
\end{tabular}%
}
\end{table}
\vspace{-2mm}

\begin{figure}[t]
  \centering
  \begin{minipage}{0.33\linewidth}
    \centering
    \includegraphics[width=\linewidth]{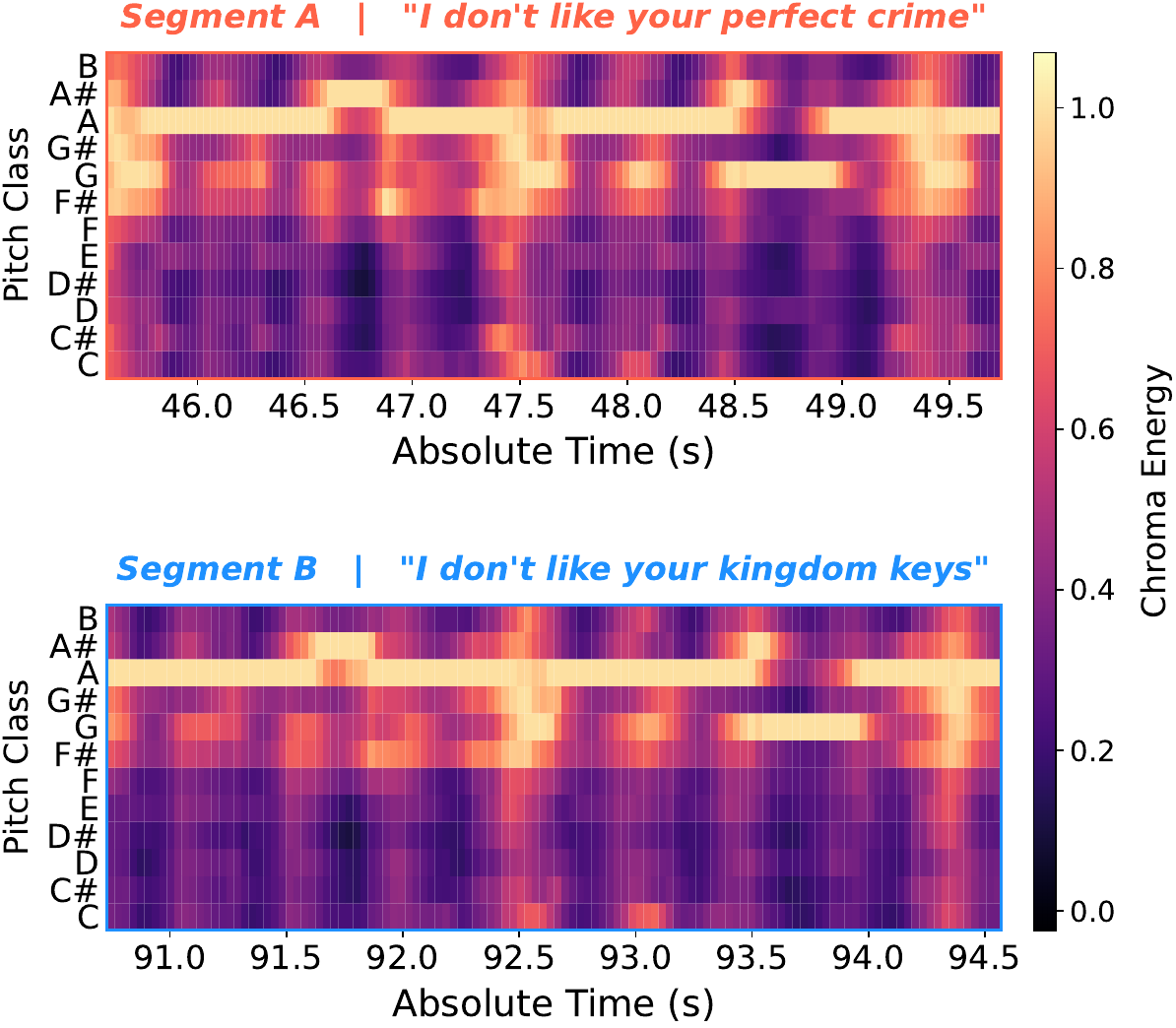}
  \end{minipage}
  \hfill 
  \begin{minipage}{0.65\linewidth}
    \centering
    \includegraphics[width=\linewidth]{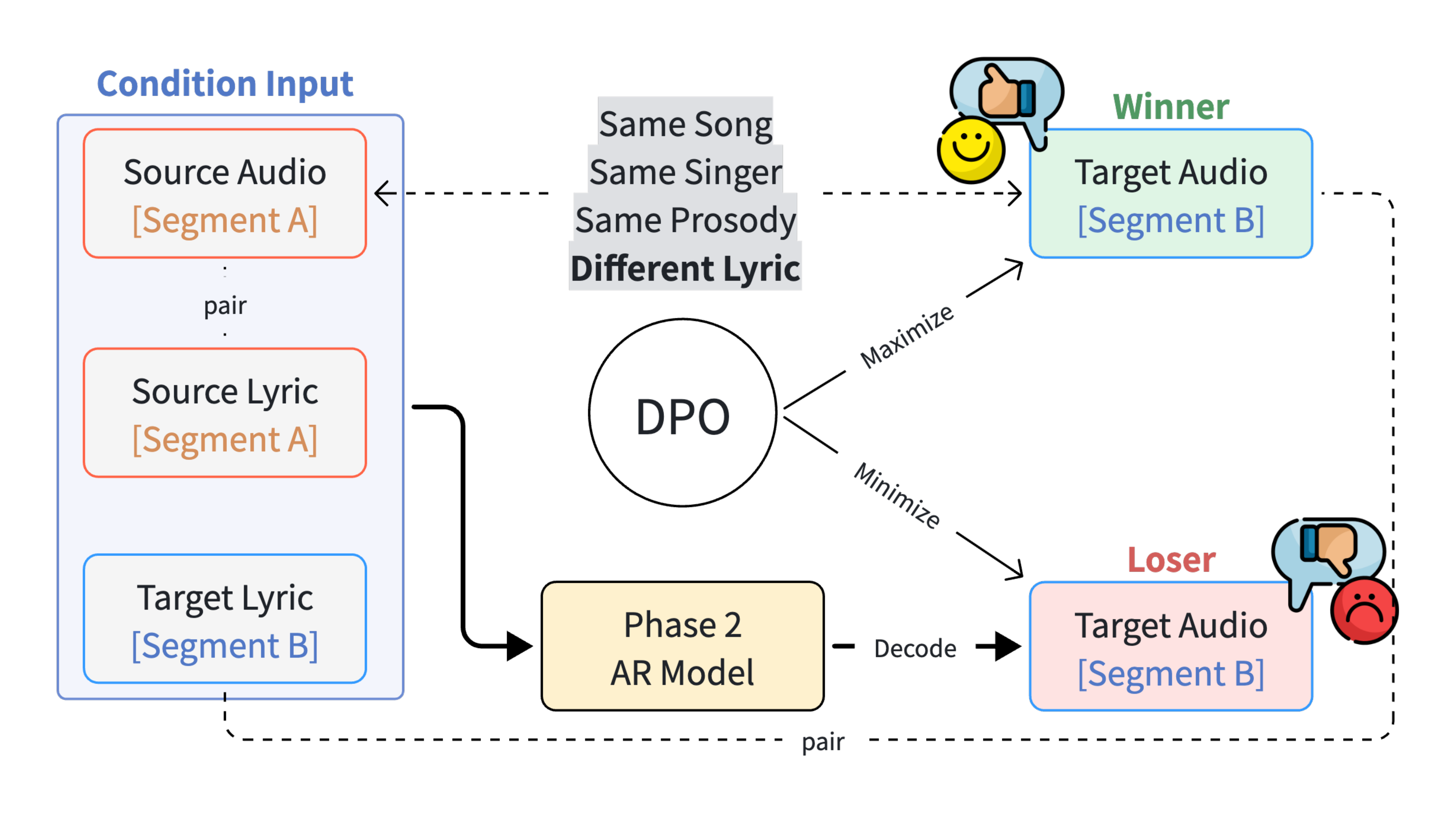}
  \end{minipage}
  \caption{Example of a Real-to-Real mined singing pair (left) and the DPO training process (right). Different lyrics (``I don't like your perfect crime'' vs.\ ``I don't like your kingdom keys'')in the same song of Look what you made me do by Taylor Swift. The song pair is only used for visualization for research purpose.}
  \label{fig:dpo_combined}
\end{figure}

To overcome the limitations of synthetic targets, our goal is to acquire Real-to-Real paired data. We draw inspiration from the inherent compositional structure of popular music, which frequently features repeated sections—such as different verses—where the same singer performs an identical melody with varying lyrics. Leveraging this property, we automatically mine pairs of real vocal segments from the same song that share the identical singer, timbre, and melody, but differ only in textual content. Since both the source and target in these pairs are natural recordings, they better match the target distribution of practical singing editing. The specific extraction and alignment pipeline is detailed as follows:

\textbf{Segment Extraction.} Vocal segments are extracted using structural metadata (e.g., verse and chorus tags) and utterance-level timestamps available in music streaming corpora, discarding segments shorter than 2.0 seconds. This structural alignment eliminates the need for external music segmentation models while providing scalable paired data. Given a song with timestamped lyrics, we extract vocal segments $S = \{s_1, s_2, \dots, s_N\}$ and discard segments shorter than 2.0 seconds.

\textbf{Feature Alignment via DTW.}
For each pair $(s_i, s_j)$ from the same song, we extract Chroma CQT features and compute a normalized DTW distance $D(s_i, s_j) = \text{DTW}(\text{CQT}(s_i), \text{CQT}(s_j)) / \max(|s_i|, |s_j|)$, where $|s|$ denotes the sequence length.

\textbf{Filtering Criteria.}
A pair is retained only if: (1) Lyrical Difference: $T_i \neq T_j$; (2) Duration Consistency: duration ratio difference within $30\%$; (3) Melodic Similarity: normalized DTW distance below threshold $\tau < 0.025$.

During DPO training, we use $s_i$ as the source condition and $T_j$ as the target text, while the real segment $s_j$ provides the preferred response $y^+$. The rejected response $y^-$ is generated by the Phase 2 AR model. We optimize the policy model $\pi_\theta$ against a frozen reference model $\pi_{\text{ref}}$:
\begin{equation}
\mathcal{L}_{\text{DPO}} = -\log \sigma \Big( \beta \big[ \log \frac{\pi_\theta(y^+|x)}{\pi_{\text{ref}}(y^+|x)} - \log \frac{\pi_\theta(y^-|x)}{\pi_{\text{ref}}(y^-|x)} \big] \Big).
\end{equation}
We construct more than 300 hours of English and about 250 hours of Chinese Real-to-Real singing pairs, corresponding to roughly one-tenth of the original song collection.

\subsection{Flow-Matching Acoustic Model}
The second stage utilizes a Flow Matching Transformer (FMT)~\cite{lipman2022flow} to convert discrete tokens predicted by the AR model into continuous mel-spectrograms. We formulate the acoustic rendering process as a conditional flow-matching problem, learning a continuous-time probability path $p_t(\mathbf{x})$ from a Gaussian prior to the target data distribution. We provide more details about flow matching in Appendix~\ref{app:fmt}.

The FMT is conditioned on the target content-style tokens, the prompt audio's content-style tokens, and the prompt's mel-spectrogram via bidirectional self-attention. To ensure high-fidelity vocal rendering while maintaining acoustic-scene continuity, we employ \textit{Mixed-Condition Training}: during training, the target mel-spectrogram is clean in 50\% of batches and corrupted with background acoustics in the remaining 50\%, while the conditioning tokens are consistently extracted from clean vocals. This conditioning asymmetry forces the acoustic model to reconstruct environmental noise and reverberation purely from the continuous mel prompt, rather than relying on discrete semantic shortcuts.

To construct noisy training conditions, we mixed clean vocals with background audio from DNS Challenge~\cite{reddy2020dns}, WHAM!~\cite{wichern2019wham}, and AudioSet~\cite{gemmeke2017audioset}, following the degradation pipeline of AnyEnhance~\cite{zhang2025anyenhance}. The SNR was uniformly sampled from $[-5, 20]$~dB, and reverberation with RT60 up to 1.5~s was simulated using synthetic and real room impulse responses (RIRs).

\section{Experiments}
\label{sec:exp}
We first describe the experimental setup, benchmark, and evaluation metrics (Section~\ref{sec:exp-set}). We then present the unified speech and singing editing results of \tool in Section~\ref{sec:main_res}, focusing on its generalization to real-world scenarios (Table~\ref{tab:main_real}), performance across diverse simulated acoustics (Table~\ref{tab:main_simulated}), and robustness under extreme acoustic conditions (Table~\ref{tab:comparison_in_extreme_conditions}). Results on clean audio are in Appendix~\ref{app:clean_result} Finally, we conduct ablation studies on our core designs (Section~\ref{sec:ablation}, Table~\ref{tab:comprehensive_ablation}) and report subjective evaluation results (Table~\ref{tab:subjective_eval}).

\subsection{Experimental Settings}
\label{sec:exp-set}

\textbf{Baselines}\quad We compare \tool with a wide range of baselines: \textbf{1) Speech editing models}: SSR-Speech~\cite{wang2025ssr}, VoiceCraft~\cite{peng2024voicecraft} (evaluated on English only), and Step-Audio-EditX~\cite{yan2025step}; \textbf{2) Singing voice editing models}: StyleSinger~\cite{zhang2024stylesinger}, TCSinger~\cite{zhang2024tcsinger}, and SoulX-Singer~\cite{qian2026soulx} (evaluated on Chinese only); \textbf{3) Unified generation models}: Ming-UniAudio~\cite{yan2025ming}, Mimo-Audio~\cite{zhang2025mimo}, and Vevo2~\cite{zhang2025vevo2}. Notably, Vevo2 is the only baseline addressing both speech and singing voice modalities, serving as our primary unified reference.

\textbf{Benchmark}\quad To rigorously assess generation and generalization capabilities, we evaluate on \toolb, a benchmark we constructed covering clean, simulated, and real-world speech and singing editing in both English and Chinese. It is organized into 12 subsets (3 subsets $\times$ 2 modalities $\times$ 2 languages), with 50 samples per subset, totaling 600 evaluation items. Full construction details are provided in Appendix~\ref{app:bench}.

\textbf{Metrics}\quad We evaluate our approach from multiple objective and subjective perspectives: \textbf{1) Content accuracy} via WER/CER using Whisper-large-v3~\cite{radford2023whisper}; \textbf{2) Speaker similarity} (SIM-spk) via WavLM cosine similarity; \textbf{3) Melodic fidelity }  for singing via log-F0 Pearson correlation (F0 Corr); \textbf{4) Acoustic environment fidelity} via CLAP background similarity~\cite{wu2023large} (SIM-bg), RT60Error, and SNRError. Subjective evaluation includes N-MOS (naturalness), S-MOS (similarity), and Fooling Rate. Full metric definitions are in Appendix~\ref{app:bench_metrics}.
\vspace{-2mm}

\subsection{Main Results}
\label{sec:main_res}

To demonstrate the superior generalization of \tool, we focus our main analysis on in-the-wild recordings and complex acoustic simulations. 

\textbf{1) Real-World Generalization}\quad As shown in Table~\ref{tab:main_real}, \tool demonstrates state-of-the-art robustness on in-the-wild real-word recordings. For speech, \tool achieves the lowest English WER (0.0616) and highest SIM-spk (0.9675). More importantly, it retains a decisive background preservation advantage over the SOTA Step-Audio-EditX (English SIM-bg: 0.7705 vs. 0.6418). This confirms that avoiding source-separation pitfalls translates to tangible acoustic continuity in unpredictable wild environments. For real-world singing, \tool comprehensively dominates in text accuracy (Chinese CER: 0.0587 vs. 0.8919 for SoulX-Singer) and melodic fidelity, confirming robust lyrical control regardless of complex background music presence.

\textbf{2) Performance in Simulated Acoustics}\quad Table~\ref{tab:main_simulated} summarizes results averaged over 15 simulated conditions. Under additive noise and reverberation, the structural advantages of \tool become prominent. Compared to Step-Audio-EditX, \tool exhibits consistently higher background continuity (e.g., Chinese SIM-bg: 0.8374 vs. 0.4896). This advantage stems directly from our separation-free architecture, avoiding mechanical artifacts and phase distortions. For singing, \tool constantly outperforms Vevo2 across nearly all metrics, validating that our simulated training paradigm generalizes robustly across diverse acoustic conditions.

\textbf{3) Robustness in Extreme Conditions}\quad To further push the limits of generalization, we compare \tool with Vevo2 under severe acoustic conditions (Table~\ref{tab:comparison_in_extreme_conditions}). Even at an extreme SNR of 0~dB or heavy reverberation (RT60=1.5~s), \tool achieves state-of-the-art robustness overall. It reduces SNR estimation errors, from 13.321 to 3.368 in speech SNR=0~dB and maintains high speaker similarity and background consistency in scenarios where Vevo2 struggles to generate coherent audio. Additionally, the clean subset results, showing \tool surpasses Vevo2 without being compromised by environment-aware training, are detailed in 
Appendix~\ref{app:clean_result}. Note that while SSR-Speech occasionally shows competitive objective scores (e.g., in clean metrics, detailed in Appendix~\ref{app:clean_result}), it operates essentially as a full-utterance text-to-speech model rather than a localized editor, fundamentally altering unedited acoustic characteristics.

\vspace{-2mm}

\subsection{Ablation Study}
\label{sec:ablation}
In Table~\ref{tab:comprehensive_ablation}, we explore the impact of several core designs of \tool, evaluated on a mixed speech-and-singing subset at 10~dB SNR.
\textbf{1) Noise-Robust Tokenizer:} Replacing this module with a standard VQ tokenizer trained only on clean data severely degrades intelligibility (WER jumps from 0.042 to 0.145). This confirms its necessity for extracting stable semantic features under noisy conditions. As analyzed in Appendix~\ref{app:distillation_ablation}, our teacher-student distillation achieves significantly higher token match rates across all SNR levels compared to direct continuous spectral supervision, ensuring reliable semantic conditioning under adverse acoustic environments. \textbf{2) Training Data Modality:} Relying exclusively on speech data destroys the model's ability to render melodies (F0 correlation drops to 0.181). Conversely, training only on singing data worsens pronunciation accuracy. This proves that our unified training strategy is essential for cross-domain generation. \textbf{3) Staged AR Training:} Omitting Phase 1 pre-training causes near-total model collapse in background similarity (SIM-bg: 0.378) and melodic fidelity. Skipping Phase 2 heavily degrades textual alignment. Furthermore, removing unmodified-area-aware supervision mechanism reduces the model's ability to perfectly preserve the original prosody in unedited regions. Detailed results on the Unmodified Area Fidelity, demonstrating how this mechanism effectively prevents prosody drift and preserves original intonation, are provided in Appendix \ref{app:unmodified_fidelity}. \textbf{4) Preference Alignment:} Removing DPO alignment and replacing our mined Real-to-Real pairs with synthesized targets yield similarly poor results. The no-DPO variant suffers from a clear drop in melodic fidelity, while the synthetic-DPO variant further degrades speech quality and intelligibility. This near-equivalence suggests that the acoustic artifacts in synthesized speech mislead preference optimization, rendering DPO almost as ineffective as having no alignment at all. It strongly supports our hypothesis that synthetic targets introduce detrimental artifacts, making natural human vocal pairs indispensable for effective preference learning. \textbf{5) Noisy Acoustic Rendering:} Training the Flow Matching Transformer solely on clean targets causes a failure in background consistency. This highlights that simulated noisy rendering is indispensable for seamless editing.

\begin{table}[h]
\centering
\caption{\textbf{Objective results on the \toolb real-world subset} English uses WER, Chinese uses CER; F0 Corr is singing-only. Bold indicates best per domain; underlined entries indicate that \tool outperforms the unified baseline Vevo2~\cite{zhang2025vevo2}. }
\label{tab:main_real}
\resizebox{\textwidth}{!}{%
\begin{tabular}{l ccccc >{\columncolor{lightgray}}c ccccc >{\columncolor{lightgray}}c}
\toprule
& \multicolumn{6}{c}{\textbf{Speech Editing}} & \multicolumn{6}{c}{\textbf{Singing Voice Editing}} \\
\cmidrule(r){2-7}\cmidrule(l){8-13}
\textbf{Metric}
  & \makecell{SSR-\\Speech} & \makecell{Voice-\\Craft} & \makecell{Step-\\EditX} & Vevo2 & \makecell{Ming-\\Uni} & \textbf{\tool}
  & \makecell{Style-\\Singer} & TCSinger & \makecell{SoulX-\\Singer} & Vevo2 & \makecell{Ming-\\Uni} & \textbf{\tool} \\
\midrule
WER (en)$\downarrow$
  & 0.0827 & 0.1081 & 0.1706 & 0.0778 & 0.5385 & \textbf{\underline{0.0616}}
  & {--} & 0.9971 & {--} & 0.2809 & 1.0000 & \textbf{\underline{0.1749}} \\
CER (zh)$\downarrow$
  & 0.1816 & {--} & \textbf{0.1296} & 0.1660 & 0.3000 & \underline{0.1433}
  & 5.4133 & 0.3171 & 0.8919 & 0.1780 & 0.1538 & \textbf{\underline{0.0587}} \\
SIM-spk (en)$\uparrow$
  & 0.9652 & 0.9604 & 0.9369 & 0.9650 & 0.9632 & \textbf{\underline{0.9675}}
  & {--} & 0.7082 & {--} & 0.8785 & 0.3009 & \textbf{\underline{0.9167}} \\
SIM-spk (zh)$\uparrow$
  & 0.8958 & {--} & 0.9054 & 0.9027 & 0.8085 & \textbf{\underline{0.9108}}
  & 0.6745 & 0.7437 & 0.8942 & 0.8970 & \textbf{0.9502} & \underline{0.9198} \\
F0 Corr (en)$\uparrow$
  & {--} & {--} & {--} & {--} & {--} & {--}
  & {--} & 0.0217 & {--} & 0.2030 & -0.1780 & \textbf{\underline{0.2080}} \\
F0 Corr (zh)$\uparrow$
  & {--} & {--} & {--} & {--} & {--} & {--}
  & 0.0217 & 0.0328 & \textbf{0.4197} & 0.3326 & -0.0697 & \underline{0.4027} \\
SIM-bg (en)$\uparrow$
  & \textbf{0.8729} & 0.7039 & 0.6418 & 0.7030 & 0.6425 & \underline{0.7705}
  & {--} & 0.3446 & {--} & 0.5671 & 0.4241 & \textbf{\underline{0.7384}} \\
SIM-bg (zh)$\uparrow$
  & \textbf{0.8472} & {--} & 0.6459 & 0.6798 & 0.8191 & \underline{0.6818}
  & 0.3527 & 0.3367 & 0.5413 & 0.5294 & 0.6301 & \textbf{\underline{0.6963}} \\
RT60Err (en)$\downarrow$
  & 0.0276 & 0.0043 & \textbf{0.0006} & 0.0654 & 0.0009 & \underline{0.0007}
  & {--} & 0.1971 & {--} & 0.1834 & 2.9556 & \textbf{\underline{0.0256}} \\
RT60Err (zh)$\downarrow$
  & 0.0149 & {--} & 0.0072 & 0.0072 & \textbf{0.0002} & 0.0075
  & 0.0091 & \textbf{0.0006} & 0.1265 & 0.0674 & {--} & \underline{0.0094} \\
SNRErr (en)$\downarrow$
  & 3.5536 & 9.6702 & 15.5598 & 13.8125 & \textbf{0.7558} & \underline{8.8241}
  & {--} & 46.1529 & {--} & 15.0969 & 38.5435 & \textbf{\underline{6.1996}} \\
SNRErr (zh)$\downarrow$
  & 6.8708 & {--} & 17.4128 & 16.9422 & \textbf{5.9399} & \underline{12.3207}
  & 45.6697 & 40.9728 & 16.4016 & 14.0181 & 7.1816 & \textbf{\underline{5.6354}} \\
\bottomrule
\end{tabular}%
}
\end{table}
\vspace{-2mm}

\begin{table}[h]
\centering
\caption{\textbf{Objective results on the \toolb simulated subset (averaged over 15 conditions)} Layout same as Table~\ref{tab:main_real}.}
\label{tab:main_simulated}
\resizebox{\textwidth}{!}{%
\begin{tabular}{l ccccc >{\columncolor{lightgray}}c ccccc >{\columncolor{lightgray}}c}
\toprule
& \multicolumn{6}{c}{\textbf{Speech Editing}} & \multicolumn{6}{c}{\textbf{Singing Voice Editing}} \\
\cmidrule(r){2-7}\cmidrule(l){8-13}
\textbf{Metric}
  & \makecell{SSR-\\Speech} & \makecell{Voice-\\Craft} & \makecell{Step-\\EditX} & Vevo2 & \makecell{Ming-\\Uni} & \textbf{\tool}
  & \makecell{Style-\\Singer} & TCSinger & \makecell{SoulX-\\Singer} & Vevo2 & \makecell{Ming-\\Uni} & \textbf{\tool} \\
\midrule
WER (en)$\downarrow$
  & 0.1085 & 0.2173 & 0.1618 & 0.0686 & \textbf{0.0588} & 0.0770
  & {--} & 1.2480 & {--} & 0.2579 & 0.5500 & \textbf{\underline{0.2368}} \\
CER (zh)$\downarrow$
  & 1.0317 & {--} & \textbf{0.0875} & 0.1031 & 0.1163 & 0.1184
  & 0.7595 & 0.6019 & 0.9028 & 0.3065 & 0.6250 & \textbf{\underline{0.2975}} \\
SIM-spk (en)$\uparrow$
  & \textbf{0.9660} & 0.9383 & 0.9363 & 0.9296 & 0.9625 & \underline{0.9535}
  & {--} & 0.5530 & {--} & 0.9015 & 0.8395 & \textbf{\underline{0.9223}} \\
SIM-spk (zh)$\uparrow$
  & 0.9056 & {--} & 0.8840 & 0.9164 & 0.9136 & \textbf{\underline{0.9328}}
  & 0.7093 & 0.7394 & 0.8723 & 0.8838 & 0.7996 & \textbf{\underline{0.9014}} \\
F0 Corr (en)$\uparrow$
  & {--} & {--} & {--} & {--} & {--} & {--}
  & {--} & -0.0221 & {--} & 0.2932 & 0.2436 & \textbf{\underline{0.3014}} \\
F0 Corr (zh)$\uparrow$
  & {--} & {--} & {--} & {--} & {--} & {--}
  & 0.0047 & 0.0135 & \textbf{0.4363} & 0.3152 & 0.0189 & \underline{0.3281} \\
SIM-bg (en)$\uparrow$
  & \textbf{0.8409} & 0.6808 & 0.5059 & 0.5445 & 0.8292 & \underline{0.7787}
  & {--} & 0.3301 & {--} & 0.5943 & 0.2380 & \textbf{\underline{0.8190}} \\
SIM-bg (zh)$\uparrow$
  & 0.8187 & {--} & 0.4896 & 0.6760 & 0.7379 & \textbf{\underline{0.8374}}
  & 0.4439 & 0.4436 & 0.6997 & 0.6793 & 0.4068 & \textbf{\underline{0.8341}} \\
RT60Err (en)$\downarrow$
  & 0.1204 & 0.0008 & \textbf{0.0006} & 0.0142 & 0.0007 & \underline{0.0021}
  & {--} & 0.0436 & {--} & 0.0262 & {--} & \textbf{\underline{0.0256}} \\
RT60Err (zh)$\downarrow$
  & 0.4526 & {--} & \textbf{0.0854} & 0.0908 & 0.0952 & 0.1637
  & 0.0533 & 0.0007 & 0.1626 & 0.1398 & {--} & \textbf{\underline{0.0004}} \\
SNRErr (en)$\downarrow$
  & \textbf{1.7351} & 5.9261 & 20.8958 & 12.8442 & 2.4145 & \underline{5.1518}
  & {--} & 38.1293 & {--} & 8.3918 & 6.0304 & \textbf{\underline{3.1332}} \\
SNRErr (zh)$\downarrow$
  & 3.7138 & {--} & 19.1547 & 7.4499 & 26.3580 & \textbf{\underline{1.7669}}
  & 38.9281 & 33.2150 & 7.5001 & 8.8160 & \textbf{3.2311} & \underline{4.1795} \\
\bottomrule
\end{tabular}%
}
\end{table}

\begin{table}[h]
\centering
\caption{\textbf{Comparison of \tool and Vevo2~\cite{zhang2025vevo2} under extreme conditions} Bold indicates the better result in each pairwise comparison. Underlined entries highlight where \tool outperforms Vevo2.}
\label{tab:comparison_in_extreme_conditions}
\resizebox{\textwidth}{!}{%
\begin{tabular}{lllccccccc}
\toprule
\textbf{Task} & \textbf{Condition} & \textbf{Model} & \textbf{SNRErr}$\downarrow$ & \textbf{RT60Err}$\downarrow$ & \textbf{WER/CER}$\downarrow$ & \textbf{SIM-spk}$\uparrow$ & \textbf{SIM-bg}$\uparrow$ & \textbf{N-MOS}$\uparrow$ & \textbf{S-MOS}$\uparrow$ \\
\midrule
\multirow{6}{*}{Speech} & \multirow{2}{*}{SNR=0 dB}
  & Vevo2 & 13.321 & \textbf{0.011} & \textbf{0.196} & 0.806 & 0.471 & 4.02 & 3.96 \\
  & & \cellcolor{lightgray}\textbf{\tool} & \cellcolor{lightgray}\textbf{\underline{3.368}} & \cellcolor{lightgray}0.165 & \cellcolor{lightgray}0.386 & \cellcolor{lightgray}\textbf{\underline{0.897}} & \cellcolor{lightgray}\textbf{\underline{0.756}} & \cellcolor{lightgray}\textbf{\underline{4.31}} & \cellcolor{lightgray}\textbf{\underline{4.27}} \\
\cmidrule(lr){2-10}
  & \multirow{2}{*}{RT60=1.5 s}
  & Vevo2 & 10.383 & 0.031 & 0.093 & 0.929 & 0.504 & 4.08 & 4.01 \\
  & & \cellcolor{lightgray}\textbf{\tool} & \cellcolor{lightgray}\textbf{\underline{1.480}} & \cellcolor{lightgray}\textbf{\underline{0.0003}} & \cellcolor{lightgray}\textbf{\underline{0.067}} & \cellcolor{lightgray}\textbf{\underline{0.940}} & \cellcolor{lightgray}\textbf{\underline{0.820}} & \cellcolor{lightgray}\textbf{\underline{4.36}} & \cellcolor{lightgray}\textbf{\underline{4.33}} \\
\cmidrule(lr){2-10}
  & \multirow{2}{*}{Both (SNR0+RT60)}
  & Vevo2 & 9.421 & \textbf{0.0005} & 0.156 & 0.884 & 0.458 & 3.95 & 3.90 \\
  & & \cellcolor{lightgray}\textbf{\tool} & \cellcolor{lightgray}\textbf{\underline{2.017}} & \cellcolor{lightgray}0.0006 & \cellcolor{lightgray}\textbf{\underline{0.105}} & \cellcolor{lightgray}\textbf{\underline{0.911}} & \cellcolor{lightgray}\textbf{\underline{0.848}} & \cellcolor{lightgray}\textbf{\underline{4.24}} & \cellcolor{lightgray}\textbf{\underline{4.18}} \\
\midrule
\multirow{6}{*}{Singing} & \multirow{2}{*}{SNR=0 dB}
  & Vevo2 & 9.940 & 0.146 & 0.530 & 0.829 & \textbf{0.783} & 4.06 & 3.98 \\
  & & \cellcolor{lightgray}\textbf{\tool} & \cellcolor{lightgray}\textbf{\underline{3.490}} & \cellcolor{lightgray}\textbf{\underline{0.017}} & \cellcolor{lightgray}\textbf{\underline{0.323}} & \cellcolor{lightgray}\textbf{\underline{0.864}} & \cellcolor{lightgray}0.205 & \cellcolor{lightgray}\textbf{\underline{4.34}} & \cellcolor{lightgray}\textbf{\underline{4.29}} \\
\cmidrule(lr){2-10}
  & \multirow{2}{*}{RT60=1.5 s}
  & Vevo2 & 6.229 & 0.117 & 0.416 & 0.893 & 0.679 & 4.11 & 4.04 \\
  & & \cellcolor{lightgray}\textbf{\tool} & \cellcolor{lightgray}\textbf{\underline{2.483}} & \cellcolor{lightgray}\textbf{\underline{0.0003}} & \cellcolor{lightgray}\textbf{\underline{0.142}} & \cellcolor{lightgray}\textbf{\underline{0.932}} & \cellcolor{lightgray}\textbf{\underline{0.850}} & \cellcolor{lightgray}\textbf{\underline{4.39}} & \cellcolor{lightgray}\textbf{\underline{4.35}} \\
\cmidrule(lr){2-10}
  & \multirow{2}{*}{Both (SNR0+RT60)}
  & Vevo2 & 7.296 & \textbf{0.0002} & 1.154 & 0.821 & 0.573 & 3.92 & 3.86 \\
  & & \cellcolor{lightgray}\textbf{\tool} & \cellcolor{lightgray}\textbf{\underline{1.809}} & \cellcolor{lightgray}0.0006 & \cellcolor{lightgray}\textbf{\underline{0.262}} & \cellcolor{lightgray}\textbf{\underline{0.904}} & \cellcolor{lightgray}\textbf{\underline{0.840}} & \cellcolor{lightgray}\textbf{\underline{4.21}} & \cellcolor{lightgray}\textbf{\underline{4.16}} \\
\bottomrule
\end{tabular}%
}
\end{table}

\begin{table}[h]
    
    \begin{minipage}[t]{0.63\textwidth}
        \vspace{0pt}
        \caption{\textbf{Ablation study on unified editing under 10dB SNR} F0 Corr is computed specifically on singing samples.}
        \label{tab:comprehensive_ablation}
        \resizebox{\linewidth}{!}{%
        \begin{tabular}{l cccccc}
        \toprule
        \textbf{Model Variation} & \textbf{WER/CER$\downarrow$} & \textbf{SIM-s$\uparrow$} & \textbf{F0 Corr$\uparrow$} & \textbf{SIM-bg$\uparrow$} & \textbf{N-MOS$\uparrow$} & \textbf{S-MOS$\uparrow$} \\
        \midrule
        \rowcolor{lightgray}
        \textbf{\tool (Full Model)} & \textbf{0.042} & \textbf{0.970} & \textbf{0.534} & \textbf{0.915} & \textbf{4.28} & \textbf{4.22} \\
        \midrule
        \rowcolor{lightgray!40}
        \multicolumn{7}{c}{\textbf{Impact of Noise-Robust Tokenizer}} \\
        \midrule
        w/o Noise-Robust Tokenizer & 0.145 & 0.900 & 0.412 & 0.720 & 3.08 & 4.18 \\
        \midrule
        \rowcolor{lightgray!40}
        \multicolumn{7}{c}{\textbf{Impact of AR Training Data Modality}} \\
        \midrule
        Speech Data Only & 0.048 & 0.962 & 0.181 & 0.905 & 4.05 & 3.52 \\
        Singing Data Only & 0.068 & 0.940 & 0.232 & 0.895 & 3.95 & 3.32 \\
        \midrule
        \rowcolor{lightgray!40}
        \multicolumn{7}{c}{\textbf{Impact of AR Training Process}} \\
        \midrule
        w/o Phase 1 & 0.049 & 0.961 & 0.042 & 0.378 & 3.21 & 3.39 \\
        w/o Phase 2 & 0.082 & 0.935 & 0.512 & 0.900 & 3.75 & 3.98 \\
        w/o Unmod.-Area-Aware Sup. & 0.055 & 0.960 & 0.403 & 0.910 & 3.98 & 3.99 \\
        w/o Preference Alignment  & 0.055 & 0.958 & 0.421 & 0.912 & 3.78 & 3.98 \\
        w/ SVS Synth Alignment Data & 0.057 & 0.959 & 0.414 & 0.913 & 3.76 & 4.01 \\
        \midrule
        \rowcolor{lightgray!40}
        \multicolumn{7}{c}{\textbf{Impact of FMT Training Data}} \\
        \midrule
        w/o Noisy Data Rendering & 0.043 & 0.965 & 0.530 & 0.200 & 3.88 & 3.51 \\
        \bottomrule
        \end{tabular}%
        }
        \end{minipage}
        \hfill
        \centering
        \begin{minipage}[t]{0.35\textwidth}
        \vspace{0pt} 
        \caption{\textbf{Subjective evaluation scores} We randomly choose 60 samples from \toolb. Comparisons can also be found on our demo page. More detailed statistics can be found in Table ~\ref{tab:subjective_detailed}}.
        \vspace{2mm} 
        \label{tab:subjective_eval}
        \resizebox{\linewidth}{!}{%
        \begin{tabular}{ll ccc}
        \toprule
        \textbf{Domain} & \textbf{Method} & \textbf{N-MOS$\uparrow$} & \textbf{S-MOS$\uparrow$} & \textbf{Fool Rate} \\
        \midrule
        \multirow{5}{*}{\textbf{Speech}}
        & Vevo2 & 4.15 & 4.02 & 31.5\% \\
        & SSR-Speech & 4.48 & 3.99 & 43.2\% \\
        & VoiceCraft & 4.04 & 4.32 & 35.7\% \\
        & Step-EditX & 3.95 & 4.02 & 33.8\% \\
        & \cellcolor{lightgray}\textbf{\tool} & \cellcolor{lightgray}\textbf{4.48} & \cellcolor{lightgray}\textbf{4.45} & \cellcolor{lightgray}\textbf{46.5\%} \\
        \midrule
        \multirow{3}{*}{\textbf{Singing}}
        & Vevo2 & 4.22 & 4.15 & 28.6\% \\
        & SoulX-singer & 3.51 & 3.34 & 0.0\% \\
        & \cellcolor{lightgray}\textbf{\tool} & \cellcolor{lightgray}\textbf{4.52} & \cellcolor{lightgray}\textbf{4.48} & \cellcolor{lightgray}\textbf{42.8\%} \\
        \bottomrule
        \end{tabular}%
        }
    \end{minipage}
\end{table}

\subsection{Case Study}
\begin{figure}[h]
  \centering
  \includegraphics[width=0.85\linewidth]{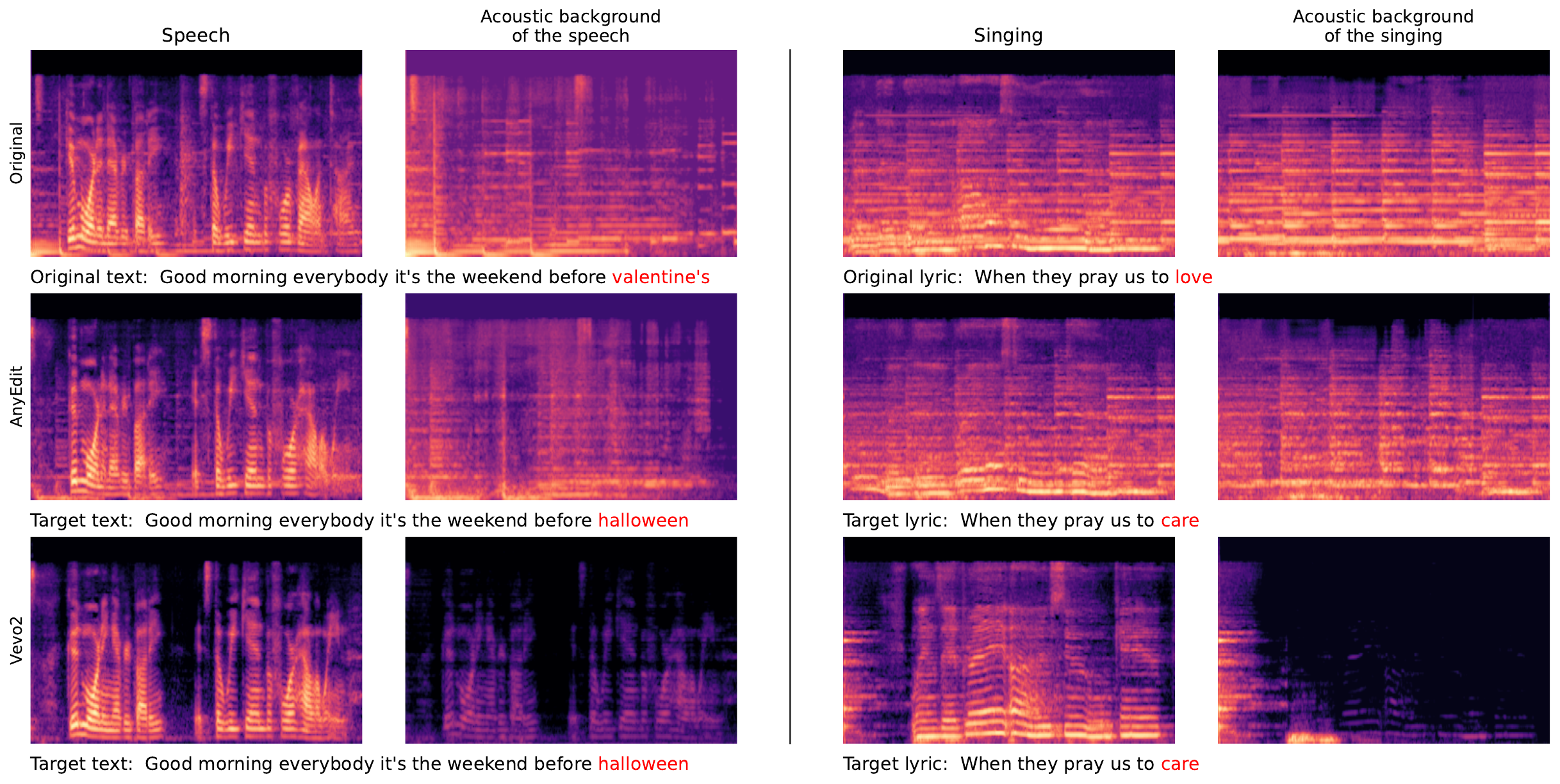}
  \caption{Case study on real-world speech and singing voice editing.}
  \label{fig:case_study}
\end{figure}

We visualize representative real-world speech and singing editing cases via Mel-spectrograms in Figure~\ref{fig:case_study}. When applying the baseline Vevo2 directly (Row 3), the background texture is severely discarded during the editing window, manifesting as stark, dark gaps in the background spectrograms. In contrast, \tool (Row 2)  modifies the target text and  remains practically identical to the original background spectrograms (Row 1). By natively modeling the unified acoustic scene, \tool bypasses the artifacts bottleneck of separation models, ensuring background continuity and realism. To further systematically compare our approach with explicit ``separation-edit-mix'' pipelines, including direct mixing approaches analogous to SeamlessEdit~\cite{chen2025seamlessedit} and SNR-aware mixing methods akin to DAIEN-TTS~\cite{lu2025daien}, we provide a comprehensive quantitative separation comparison in Appendix \ref{app:separation_comparison}

\subsection{Subjective Evaluation}

As shown in Table~\ref{tab:subjective_eval}, we conducted a subjective evaluation with 10 raters. Compared with Vevo2, \tool yields consistently higher N-MOS and S-MOS across both domains. Notably, \tool achieves significant gains in Fooling Rate (where 50\% indicates perfect perceptual indistinguishability from real audio). This demonstrates that listeners much more frequently perceived our edited outputs as authentic, unedited natural recordings.

\section{Conclusion}
In this paper, we presented \tool, a unified, separation-free framework for speech and singing voice editing in real-world acoustic environments. By disentangling clean-space semantic editing from context-aware acoustic rendering, \tool overcomes the limitations of traditional direct end-to-end modeling and separation-edit-mix pipelines. Our two-stage architecture leverages an instruction-guided autoregressive editor and a flow-matching acoustic model, supported by a noise-robust tokenizer for stable prosody extraction. Additionally, unmodified-area-aware supervision and DPO on mined Real-to-Real singing pairs enhance prosodic stability and naturalness.

Furthermore, we introduced \toolb, the first comprehensive benchmark for speech and singing editing across diverse acoustic conditions. Extensive evaluations confirmed that \tool achieves state-of-the-art performance, maintaining high content accuracy, timbre fidelity, and acoustic continuity under clean, simulated, and extreme real-world degradations.

\begin{ack}
The work of Zhizheng Wu is supported by the National Natural Science Foundation of China under Grant 62376237.
This work is supported by the Program for Guangdong Introducing Innovative and Entrepreneurial Teams, Grant No. 2023ZT10X044.
\end{ack}

\bibliographystyle{unsrtnat}
\bibliography{ref_used}

\appendix

\section{Related Work}
\label{app:related}

\subsection{Speech Editing}
Speech editing has evolved from waveform concatenation and mask-based reconstruction to neural generative infilling. Early work such as \textit{CampNet}~\cite{wang2022campnet} framed editing as a localized text-to-speech (TTS) task with context-aware mask prediction. \textit{EditSpeech}~\cite{tan2021editspeech} introduced bidirectional context fusion, while \textit{VoiceCraft}~\cite{peng2024voicecraft} leveraged token rearrangement in neural codec language models to enable autoregressive bidirectional infilling. \textit{FluentSpeech}~\cite{jiang2023fluentspeech} further explored diffusion-based editing, and \textit{SSR-Speech}~\cite{wang2025ssr} improved robustness and stability in zero-shot text-based speech editing. More recent systems such as \textit{Step-Audio-EditX}~\cite{yan2025step} continued to extend audio editing capabilities. However, most existing methods were primarily developed for clean speech and did not explicitly preserve background acoustics in real-world recordings.

\subsection{Singing Voice Editing and Synthesis}
Singing voice editing is more challenging because lyric replacement must remain consistent with melody, rhythm, and vocal expression. \textit{EditSinger}~\cite{zhang2022editsinger} addressed this by disentangling pitch from lyrics to preserve melodic contours. \textit{StyleSinger}~\cite{zhang2024stylesinger}, \textit{TCSinger}~\cite{zhang2024tcsinger}, and \textit{SoulX-Singer}~\cite{qian2026soulx} demonstrated recent progress in controllable or zero-shot singing voice synthesis (SVS), while \textit{DiTSinger}~\cite{xu2025ditsinger} and \textit{SongEditor}~\cite{yang2025songeditor} explored scalable singing generation and editing via implicit alignment or dual-track masking. Nevertheless, these approaches mainly targeted clean-condition synthesis or editing and typically ignored background-scene continuity. Furthermore, many singing editing pipelines relied on synthetic target construction, inadvertently introducing artifacts from the synthesis model.

\subsection{Unified Speech and Singing Generation}
Unified modeling across speech and singing enables shared representation learning across both modalities. \textit{Vevo}~\cite{zhang2025vevo} and \textit{Vevo2}~\cite{zhang2025vevo2} showed that explicit prosody tokenization was effective for unified clean-condition generation and controllable rendering. However, these models were designed for generation rather than localized vocal editing, relied on externally provided prosodic conditions, and did not explicitly address acoustic scene preservation under noise, reverberation, or background music.

\subsection{Environment-Aware and Unified Audio Generation}
\textit{SeamlessEdit}~\cite{chen2025seamlessedit} followed a ``separation-edit-mix'' paradigm that explicitly separated vocals from the background before editing. However, such pipelines can be fragile when edited duration changes or source separation is imperfect. \textit{DAIEN-TTS}~\cite{lu2025daien} moved toward environment-aware generation by conditioning synthesis on background representations. More general models—\textit{Voicebox}~\cite{le2023voicebox}, \textit{UniAudio}~\cite{yang2023uniaudio}, \textit{Ming-UniAudio}~\cite{yan2025ming}, and \textit{MiMo-Audio}~\cite{zhang2025mimo}—further broadened unified modeling across diverse audio tasks, yet none were specifically tailored for unified speech and singing voice editing in real-world environments.

\section{Definitions of Core Representations}
\label{sec:definitions}

To establish a clear and self-contained theoretical foundation for unified vocal editing, we factorize the continuous audio signal into three mutually orthogonal feature streams, following the disentanglement principles in~\cite{zhang2025vevo, zhang2025vevo2}:

\begin{itemize}[leftmargin=3.5ex, itemsep=0.8ex]
    \item \textbf{Prosody Tokens ($z^{\text{prosody}}$):} 
    Prosody tokens capture the temporal pitch contour (fundamental frequency $F_0$), melodic trajectory, and duration dynamics across speech and singing. They are derived from normalized chromagrams to ensure octave invariance and melodic compactness. Importantly, this representation explicitly isolates melodic movement and rhythm while discarding linguistic phonetics, expressive vocal nuances, speaker identity (timbre), and environmental background acoustics.

    \item \textbf{Content-Style Tokens ($z^{\text{CS}}$):} 
    Content-style tokens encode linguistic information (phonetic transcriptions) alongside local expressive delivery habits, such as articulation manner, emotional tone, and vocal techniques (e.g., breathiness and vibrato). While content-style tokens capture fine-grained phonetic semantics and expressive dynamics, they are strictly invariant to speaker identity (timbre) and external acoustic environments (e.g., noise, reverberation).

    \item \textbf{Acoustic Context and Rendering:} 
    Acoustic context refers to the continuous time-frequency prompt (mel-spectrogram) extracted from the unedited reference audio. It encapsulates speaker-specific timbre alongside environmental acoustic properties, including ambient noise, room impulse responses (reverberation), and background music. Correspondingly, \textit{acoustic rendering} is the generative process handled by the Flow Matching Transformer (FMT), which synthesizes the post-edit mel-spectrogram by conditioning on the generated clean semantic tokens while strictly anchoring timbre and environmental textures to the reference acoustic context.

    \item \textbf{Semantic Editing:} 
    In the scope of this work, \textit{semantic editing} is defined specifically as localized textual and linguistic modifications—namely, inserting, deleting, or substituting words or lyrics within an existing utterance—while anchoring unedited content, prosody, and background continuity. Modifying high-level expressive performance styles (e.g., changing vocal emotion or singing technique) is considered orthogonal and deferred to future work.
\end{itemize}

\begin{table}[h]
\centering
\caption{Taxonomy and orthogonal properties of core representations in \tool.}
\label{tab:feature_decoupling}
\resizebox{\textwidth}{!}{%
\begin{tabular}{l l l l}
\toprule
\textbf{Representation} & \textbf{Encoded Attributes} & \textbf{Excluded Attributes} & \textbf{System Role} \\
\midrule
\textbf{Prosody Tokens} ($z^{\text{prosody}}$) 
  & Pitch contour, melody, rhythm 
  & Phonetics, vocal style, timbre, acoustics 
  & Condition for AR editor \\
\textbf{Content-Style Tokens} ($z^{\text{CS}}$) 
  & Phonemes, articulation, vibrato, emotion 
  & Speaker timbre, room/noise acoustics 
  & Output target of AR editor \\
\textbf{Acoustic Context (Mel)} 
  & Speaker timbre, background noise, RT60 
  & Semantic text content 
  & In-context prompt for FMT \\
\bottomrule
\end{tabular}%
}
\end{table}

\noindent\textbf{Decoupling Principle across Pipeline Stages:}
This factorization defines the functional boundary between our two generative stages. In Stage~1 (AR editor), degraded audio only affects the extraction of $z^{\text{prosody}}$. By equipping the prosody tokenizer with noise robustness, the entire semantic prediction process operates purely in the clean domain, generating clean $z^{\text{CS}}$ conditioned on clean $z^{\text{prosody}}$ and target text. In Stage~2 (FMT acoustic rendering), because $z^{\text{CS}}$ carries neither speaker timbre nor background noise, the acoustic renderer is prevented from exploiting semantic shortcuts, forcing it to reconstruct the continuous acoustic scene purely from the source context.

\section{AnyEditBench and Evaluation Details}

\subsection{Benchmark Construction}
\label{app:bench}
\toolb is organized into 12 subsets (3 subsets $\times$ 2 modalities $\times$ 2 languages), with 50 samples per subset and 600 evaluation items in total. All subsets contain insertion, deletion, and substitution edits with random span lengths. Table~\ref{tab:eval_datasets} provides a detailed summary.

\begin{table}[h]
\centering
\caption{Summary of \toolb. Each subset contains 50 evaluation items. The clean subset includes both single-span and multi-span edits.}
\label{tab:eval_datasets}
\resizebox{\textwidth}{!}{%
\begin{tabular}{l l l l l c}
\toprule
\textbf{subset} & \textbf{Domain} & \textbf{Language} & \textbf{Source} & \textbf{Acoustic condition} & \textbf{Items} \\
\midrule
\multirow{4}{*}{Clean}
& Speech & EN & LibriTTS, VCTK & Clean recordings; single-span and multi-span edits & 50 \\
& Speech & ZH & AISHELL-1/2/3 & Clean recordings; single-span and multi-span edits & 50 \\
& Singing & EN & Clean SingVERSE test subset~\cite{jiang2025singverse} & Clean recordings; single-span and multi-span edits & 50 \\
& Singing & ZH & Clean SingVERSE test subset~\cite{jiang2025singverse} & Clean recordings; single-span and multi-span edits & 50 \\
\midrule
\multirow{4}{*}{Simulated}
& Speech & EN & Clean-source speech subsets & \makecell[l]{15 noise/reverberation conditions \\ (SNR $\in \{0,10,20,\infty\}$, RT60 $\in \{0,0.5,1.0,1.5\}$)} & 50 \\
& Speech & ZH & Clean-source speech subsets & \makecell[l]{15 noise/reverberation conditions (excluding clean case)} & 50 \\
& Singing & EN & Clean SingVERSE test subset~\cite{jiang2025singverse} & \makecell[l]{15 noise/reverberation conditions (excluding clean case)} & 50 \\
& Singing & ZH & Clean SingVERSE test subset~\cite{jiang2025singverse} & \makecell[l]{15 noise/reverberation conditions (excluding clean case)} & 50 \\
\midrule
\multirow{4}{*}{Real-world}
& Speech & EN & YODAS~\cite{li2023yodas} & Naturally recorded environments & 50 \\
& Speech & ZH & YODAS~\cite{li2023yodas} & Naturally recorded environments & 50 \\
& Singing & EN & SingVERSE~\cite{jiang2025singverse} & Naturally recorded environments & 50 \\
& Singing & ZH & SingVERSE~\cite{jiang2025singverse} & Naturally recorded environments & 50 \\
\bottomrule
\end{tabular}%
}
\end{table}

\noindent\textbf{Clean subset:} English speech subsets were drawn from LibriTTS~\cite{zen2019libritts} and VCTK~\cite{yamagishi2019vctk}; Chinese speech subsets used AISHELL-1/2/3~\cite{bu2017aishell,du2018aishell,AISHELL3}. Clean singing evaluation used the clean subset of SingVERSE~\cite{jiang2025singverse}. Each subset contains both single-span and multi-span examples.

\noindent\textbf{Simulated subset:} Starting from clean-source speech and singing data, we synthesized degraded evaluation samples by combining noise and reverberation conditions. The noise setting was selected from SNR $\in \{0, 10, 20, \infty\}$~dB, and the reverberation setting from RT60 $\in \{0, 0.5, 1.0, 1.5\}$~s. Excluding the fully clean+dry case ($\text{SNR}=\infty, \text{RT60}=0$) yielded 15 acoustic conditions in total. Noise and RIR resources were collected from Urgent Challenge noise data~\cite{zhang2024urgent}, FSD50K~\cite{fonseca2021fsd50k}, DESED~\cite{turpault2019sound}, and a self-collected dataset.

\noindent\textbf{Real-world subset:} Real-world speech data were taken from YODAS~\cite{li2023yodas}, and real-world singing data from SingVERSE~\cite{jiang2025singverse}. These recordings were collected in naturally occurring acoustic environments.

\subsection{Evaluation Metrics}
\label{app:bench_metrics}
\subsubsection{Objective Evaluation Metrics}
\begin{itemize}
    \item \textbf{WER / CER ($\downarrow$):} We used \textit{Whisper-large-v3} \cite{radford2023whisper} to compute English WER and Chinese CER for edited content.
    \item \textbf{SIM-spk ($\uparrow$):} We computed the speaker cosine similarity between the source and edited audio using embeddings extracted with the pretrained WavLM~\footnote{\url{https://huggingface.co/microsoft/wavlm-base-plus-sv}}.
    \item \textbf{F0 Corr ($\uparrow$):} To assess the melodic quality of singing voice editing, we resampled audio to 16~kHz, extracted F0 with librosa YIN~\footnote{\url{https://github.com/librosa/librosa}}, and computed the Pearson correlation of log-F0 over vocal frames between the edited and reference audio.
    \item \textbf{SIM-bg ($\uparrow$):} To assess the preservation of the acoustic environment, we used CLAP~\cite{wu2023large} \footnote{\url{https://huggingface.co/laion/clap-htsat-unfused}} at 48~kHz and computed the cosine similarity between the background embeddings of the reference and edited audio.
    \item \textbf{RT60Error ($\downarrow$):} We first separated each waveform into vocal and accompaniment using an optimized SingNet pipeline~\cite{gu2025singnet} (identical to the benchmark construction operation), estimated the room impulse characteristics from the separated vocal track, and then computed RT60 from the estimated decay curve. We reported the absolute RT60 error (seconds) between the edited result and the reference.
    \item \textbf{SNRError ($\downarrow$):} Following the SingNet \cite{gu2025singnet} pipeline, we used the separated vocal and background stems to compute the signal-to-noise ratio as the energy ratio between vocal and accompaniment, and reported the absolute SNR error (dB) with respect to the reference clip.
    \end{itemize}

\subsubsection{Subjective Evaluation Metrics}
    We conducted human evaluation with 10 subjects, sampling 15 items from each of the speech-English, speech-Chinese, singing-English, and singing-Chinese subsets. Subjects rated \textbf{N-MOS} for naturalness and \textbf{S-MOS} for similarity.  
    \noindent
    \textbf{N-MOS (Naturalness Mean Opinion Score)} assesses the naturalness of the audio and consists of three sub-ratings:
    \begin{itemize}
        \item Naturalness of the human voice,
        \item Naturalness of the background sound,
        \item Naturalness of the mixture between background sound and human voice.
    \end{itemize}
    
    \noindent
    \textbf{S-MOS (Similarity Mean Opinion Score)} assesses the similarity to a reference and consists of two sub-ratings:
    \begin{itemize}
        \item Similarity of the human voice,
        \item Similarity of the background sound.
    \end{itemize}
    
    We also conducted a \textbf{Fooling Rate} blind A/B test in which subjects identified the authentic human recording; a Fooling Rate closer to 50\% indicates stronger perceptual indistinguishability.


\section{Implementation Details}
\label{appendix:imp_detail}
\subsection{Computational Budget and Model Parameters}
\label{app:compute_budget}
To assess training and inference efficiency, we compare the parameter scale and language/domain coverage of \tool against existing state-of-the-art audio editing and generation models in Table~\ref{tab:param_comparison}. \tool is a unified model capable of handling both speech and singing editing across English and Chinese within a single 1.1B parameter framework.

\begin{table}[h]
\centering
\caption{Comparison of model parameter counts and capability coverage.}
\label{tab:param_comparison}
{\footnotesize
\begin{tabular}{l l c c}
\toprule
\textbf{Model} & \textbf{Supported Domain} & \textbf{Trained for Editing} & \textbf{Total for Inference} \\
\midrule
\rowcolor{lightgray}
\textbf{\tool (Ours)} & Speech + Singing (EN + ZH) & 872M & 1.1B \\
Vevo2~\cite{zhang2025vevo2} & Speech + Singing (EN + ZH) & 872M & 1.1B \\
Ming-UniAudio~\cite{yan2025ming} & Speech (EN + ZH) & $\sim$2.8B active & 18.0B ($\sim$2.8B active) \\
MiMo-Audio~\cite{zhang2025mimo} & Speech (EN + ZH) & 8.0B & 9.3B \\
Step-Audio-EditX~\cite{yan2025step} & Speech (EN + ZH) & 3.5B & 3.7B \\
SSR-Speech~\cite{wang2025ssr} & Speech (EN + ZH)$^{*}$ & 840M + 840M & 1.8B \\
VoiceCraft~\cite{peng2024voicecraft} & Speech (EN only) & 830M & 0.9B \\
SoulX-Singer~\cite{qian2026soulx} & Singing (ZH only) & 443M & 1.8B \\
TCSinger~\cite{zhang2024tcsinger} & Singing (EN + ZH) & 330M & 0.34B \\
StyleSinger~\cite{zhang2024stylesinger} & Singing (ZH only) & 64M & 0.08B \\
\bottomrule
\end{tabular}}
\vspace{1mm}

\footnotesize{$^{*}$SSR-Speech requires two separate checkpoints for English and Chinese, respectively.}
\end{table}

\paragraph{Hardware and Training Resources.} 
All models were trained using the computational resources similar to Vevo2.
\subsection{AR Model}
\label{appendix:model_arch}
All our models follow the standard Transformer~\cite{vaswani2017attention, touvron2023llama} architecture, employ RoPE positional encoding~\cite{su2024roformer} and the SiLU~\cite{elfwing2018sigmoid} activation function. The encoder and decoder of the tokenizer and MGM models use bidirectional attention, while the AR models adopt causal attention.

\tool is initialized from the textual LLM \texttt{Qwen2.5-0.5B-Instruct}~\cite{yang2024qwen2}: a 12-layer Transformer decoder with hidden size 1024, with vocabulary expanded to include prosody and content-style tokens (509M parameters total). We used the same text tokenizer as Vevo2~\cite{zhang2025vevo2}. 

\subsection{FMT Model}
\label{app:fmt}
We provide additional details of the flow matching framework used to train the acoustic generation process of \tool. Flow matching~\cite{lipman2022flow} defines a continuous transformation from a prior distribution (\textit{e.g.}, Gaussian noise) to a target data distribution (\textit{e.g.}, mel-spectrograms) by learning a time-dependent velocity field along an interpolated trajectory $\boldsymbol{x}_t$.

While multiple interpolation strategies can be used to construct $\boldsymbol{x}_t$, we adopt the \textit{optimal transport path} formulation~\cite{lipman2022flow, liu2022flow}, instantiated in this work as simple linear interpolation. Specifically, given a clean mel-spectrogram $\boldsymbol{x} \in \mathbb{R}^{T \times d}$ and a noise sample $\boldsymbol{\epsilon} \sim \mathcal{N}(\boldsymbol{0}, \boldsymbol{I})$, we construct an intermediate sample as:
\begin{equation}
\boldsymbol{x}_t = t \boldsymbol{x} + (1 - t) \boldsymbol{\epsilon}, \quad t \sim \text{Uniform}(0, 1),
\end{equation}
where $t$ is sampled uniformly from $[0, 1]$, and $\boldsymbol{x}_t$ denotes the noisy input at time $t$. The corresponding ground-truth velocity is the temporal derivative of $\boldsymbol{x}_t$:
\begin{equation}
\boldsymbol{v} = \frac{d \boldsymbol{x}_t}{dt} = \boldsymbol{x} - \boldsymbol{\epsilon}.
\end{equation}
The diffusion decoder $\mathcal{D}_\phi$ is trained to predict $\boldsymbol{v}$, conditioned on the token sequence $\boldsymbol{q} = \mathcal{Q}(\mathcal{E}_\theta(\boldsymbol{x}))$ and the associated text $\boldsymbol{x}_{text}$, using the following objective:
\begin{equation}
\mathcal{L}_{\text{diff}} = \mathbb{E}_{(\boldsymbol{x}, \boldsymbol{x}_{text}), \boldsymbol{\epsilon}, t} \left[ \left\| (\boldsymbol{x} - \boldsymbol{\epsilon}) - \mathcal{D}_\phi(\boldsymbol{q}, \boldsymbol{x}_t, t, \boldsymbol{x}_{text}) \right\| \right].
\end{equation}
\paragraph{Inference} At inference time, we start with a noise sample $\boldsymbol{x}_0 = \boldsymbol{\epsilon} \sim \mathcal{N}(\boldsymbol{0}, \boldsymbol{I})$ and solve the ordinary differential equation:
\begin{equation}
\frac{d\boldsymbol{x}_t}{dt} = \mathcal{D}_\phi(\boldsymbol{q}, \boldsymbol{x}_t, t, \boldsymbol{x}_{text})
\end{equation}
from $t = 0$ to $t = 1$ using a simple Euler ODE solver over a discretized set of $N$ time steps.

Flow matching provides a stable and interpretable training signal by directly supervising the instantaneous direction in which a noisy sample $\boldsymbol{x}_t$ should evolve to match the clean target $\boldsymbol{x}$. In our setting, it enables effective training under real-world acoustic conditions, containing both speaker timbre information and background acoustic information.

\paragraph{FMT Architecture and Conditioning.}
The Flow Matching Transformer (FMT) is a non-autoregressive Transformer. Specifically, the model is constructed by stacking LLaMA-2 Decoder blocks, with the causal mask replaced by bidirectional full attention, allowing the model to attend to the sequence globally. The conditioning mechanisms are as follows:
\begin{itemize}[leftmargin=3ex, itemsep=0ex]
    \item \textbf{Time Step Injection:} The continuous time step $t$ is encoded using sinusoidal embeddings.
    \item \textbf{Condition Injection:} Both the time embeddings and the conditioning representations are injected into the Transformer blocks using Adaptive Layer Normalization (AdaLN).
    \item \textbf{Acoustic Prompting:} The reference audio prefix (capturing the target speaker's timbre and ambient background characteristics) is provided as an acoustic prompt to guide the flow-matching process in rendering the final mel-spectrogram with consistent speaker identity and acoustic conditions.
\end{itemize}

\section{Baselines}
\label{app:baseline}

\subsection{Speech Editing Baselines}
\label{appendix:baseline_speech}

\textbf{VoiceCraft}~\cite{peng2024voicecraft}\quad
A token-infilling neural codec language model designed for zero-shot speech editing and text-to-speech in the wild. It employs causal masking and delayed stacking to enable autoregressive generation within an existing sequence. We use the official code and checkpoints\footnote{\url{https://github.com/jasonppy/VoiceCraft}}.

\textbf{SSR-Speech}~\cite{wang2025ssr}\quad
A Transformer decoder-based autoregressive neural codec model tailored for stable and robust zero-shot text-based speech editing. It incorporates frame-level watermarking and classifier-free guidance. We use the official codebase and pre-trained checkpoints\footnote{\url{https://github.com/WangHelin1997/SSR-Speech}}.

\textbf{Step-Audio-EditX}~\cite{yan2025step}\quad
A state-of-the-art large language model-based tool developed by StepFun for precise, robust, and expressive speech editing. We use the official model checkpoints available on HuggingFace\footnote{\url{https://huggingface.co/stepfun-ai/Step-Audio-EditX}}.

\subsection{Unified Audio Models}
\label{appendix:baseline_unified}

\textbf{Ming-UniAudio}~\cite{yan2025ming}\quad
A comprehensive unified audio generation model built upon a 16-billion parameter architecture. It supports a wide variety of audio-related tasks including generation and editing. We use the official pre-trained checkpoint\footnote{\url{https://huggingface.co/inclusionAI/Ming-UniAudio-16B-A3B}}.

\textbf{MiMo-Audio}~\cite{zhang2025mimo}\quad
A 7-billion parameter unified audio generation model developed by Xiaomi, designed to process and generate diverse audio content across different modalities. We utilize the official base checkpoint\footnote{\url{https://huggingface.co/XiaomiMiMo/MiMo-Audio-7B-Base}}.

\textbf{Vevo2}~\cite{zhang2025vevo2}\quad
A strong unified generative model capable of high-quality speech and singing voice generation. It acts as our primary unified baseline covering both modalities in clean conditions. We use the official implementation and checkpoints. \textbf{Vevo2}~\cite{zhang2025vevo2} serves as our primary unified baseline covering both speech and singing modalities. While Vevo2 is designed for zero-shot generation (TTS/SVS) and conversion (VC/SVC) in clean acoustic conditions, it exhibits two structural limitations in real-world localized editing: (1) it does not model or reconstruct background environmental acoustics, discarding ambient textures during editing; and (2) its autoregressive design couples sequence length with prosody preservation, making it challenging to simultaneously preserve original prosody while handling edits with variable duration. In contrast, \tool decouples semantic editing from acoustic rendering to resolve this trade-off. Furthermore, as evaluated in Appendix~\ref{app:zero_shot_ext}, \tool naturally inherits strong zero-shot VC and SVC capabilities through its acoustic renderer without task-specific tuning.

\subsection{Singing Voice Baselines}
\label{appendix:baseline_singing}

\textbf{StyleSinger}~\cite{zhang2024stylesinger}\quad
A zero-shot singing voice synthesis model designed to generate high-fidelity and expressive vocals while capturing diverse singing styles. We use the official implementation and checkpoints\footnote{\url{https://github.com/AaronZ345/StyleSinger}}.

\textbf{TCSinger}~\cite{zhang2024tcsinger}\quad
A zero-shot singing voice synthesis model that provides highly controllable text-to-singing generation and timbre preservation. We use the official open-source repository and checkpoints\footnote{\url{https://github.com/AaronZ345/TCSinger}}.

\textbf{SoulX-Singer}~\cite{qian2026soulx}\quad
A high-quality, expressive singing voice synthesis model developed by Soul AILab. 
Although the model is claimed to support lyric editing, no public code is available for this functionality. 
Therefore, based on the model's inference process, we implemented our own inference code for lyric editing. 
The base model (weights) is used from the official repository\footnote{\url{https://github.com/Soul-AILab/SoulX-Singer}}, while the lyric editing inference is our own implementation.

\section{Extended Quantitative Analysis}

\subsection{Objective results on the \toolb clean split}
\label{app:clean_result}
Speech editing (left) and singing voice editing (right) are reported jointly. English uses WER, Chinese uses CER; F0 Corr is singing-only (``--'' for speech). Bold: best per domain; underlined: \tool outperforms Vevo2~\cite{zhang2025vevo2}. VoiceCraft~\cite{peng2024voicecraft}: English only; StyleSinger~\cite{zhang2024stylesinger}, SoulX-Singer~\cite{qian2026soulx}: Chinese only. Results are showed in Table \ref{tab:main_clean}.

\subsection{Evaluation on Vevo2~\cite{zhang2025vevo2}'s editing test set.}
We tested \tool on the exact datasets used by Vevo2 for expressive speech~\cite{genshin_voice_dataset} and singing voice~\cite{singstyle111, gtsinger}editing task, which predominantly feature clean backgrounds. As shown below, \tool consistently outperforms Vevo2 across core metrics on its Vevo2~\cite{zhang2025vevo2}'s Editing testsets. Results are showed in Table \ref{tab:vevo2_editing_testset}.
\begin{table}[h]
\centering
\caption{Comparison between Vevo2 and \tool on Vevo 2's speech and singing voice editing test sets.}
\label{tab:vevo2_editing_testset}
\small
\begin{tabular}{l c >{\columncolor{lightgray}}c c >{\columncolor{lightgray}}c}
\toprule
& \multicolumn{2}{c}{\textbf{Speech Editing}} & \multicolumn{2}{c}{\textbf{Singing Voice Editing}} \\
\cmidrule(r){2-3}\cmidrule(l){4-5}
\textbf{Metric} & Vevo2 & \textbf{\tool} & Vevo2 & \textbf{\tool} \\
\midrule
WER (en)$\downarrow$     & 0.1717          & \textbf{0.1658} & 0.2795          & \textbf{0.2754} \\
CER (zh)$\downarrow$     & 0.1950          & \textbf{0.1841} & 0.1986          & \textbf{0.1883} \\
SIM-spk (en)$\uparrow$   & 0.9582          & \textbf{0.9583} & 0.9496          & \textbf{0.9519} \\
SIM-spk (zh)$\uparrow$   & 0.9582          & \textbf{0.9597} & \textbf{0.9574} & 0.9553          \\
F0 Corr (en)$\uparrow$   & 0.1801          & \textbf{0.1803} & 0.3341          & \textbf{0.3406} \\
F0 Corr (zh)$\uparrow$   & 0.1989          & \textbf{0.2079} & 0.3295          & \textbf{0.3485} \\
\bottomrule
\end{tabular}
\end{table}

\label{result:clean}
\begin{table}[h]
\centering
\caption{Objective results on the \toolb clean split}
\label{tab:main_clean}
\resizebox{\textwidth}{!}{%
\begin{tabular}{l cccccc >{\columncolor{lightgray}}c cccccc >{\columncolor{lightgray}}c}
\toprule
& \multicolumn{7}{c}{\textbf{Speech Editing}} & \multicolumn{7}{c}{\textbf{Singing Voice Editing}} \\
\cmidrule(r){2-8}\cmidrule(l){9-15}
\textbf{Metric}
  & \makecell{SSR-\\Speech} & \makecell{Voice-\\Craft} & \makecell{Step-\\EditX} & Vevo2 & \makecell{Ming-\\Uni} & \makecell{Mimo-\\Audio} & \textbf{\tool}
  & \makecell{Style-\\Singer} & TCSinger & \makecell{SoulX-\\Singer} & Vevo2 & \makecell{Ming-\\Uni} & \makecell{Mimo-\\Audio} & \textbf{\tool} \\
\midrule
WER (en)$\downarrow$
  & \textbf{0.0246} & 0.0428 & 0.0823 & 0.0600 & 0.1291 & 0.0423 & \underline{0.0408}
  & {--} & 1.6463 & {--} & 0.1482 & 0.4123 & 0.1861 & \textbf{\underline{0.1425}} \\
CER (zh)$\downarrow$
  & 0.0877 & {--} & 0.0419 & 0.0561 & 0.1091 & \textbf{0.0274} & \underline{0.0487}
  & 0.6124 & 0.3135 & 0.8572 & 0.2203 & 0.4444 & 0.1957 & \textbf{\underline{0.0613}} \\
SIM-spk (en)$\uparrow$
  & \textbf{0.9876} & 0.9714 & 0.9723 & 0.9739 & 0.9692 & 0.9448 & \underline{0.9763}
  & {--} & 0.6569 & {--} & 0.9410 & 0.8314 & 0.8604 & \textbf{\underline{0.9551}} \\
SIM-spk (zh)$\uparrow$
  & \textbf{0.9655} & {--} & 0.9430 & 0.9582 & 0.9274 & 0.9344 & 0.9540
  & 0.7865 & 0.8309 & 0.9182 & 0.9458 & 0.8444 & 0.8836 & \textbf{\underline{0.9486}} \\
F0 Corr (en)$\uparrow$
  & {--} & {--} & {--} & {--} & {--} & {--} & {--}
  & {--} & -0.0538 & {--} & 0.1338 & 0.0120 & 0.0330 & \textbf{\underline{0.1589}} \\
F0 Corr (zh)$\uparrow$
  & {--} & {--} & {--} & {--} & {--} & {--} & {--}
  & 0.0081 & 0.0391 & \textbf{0.3784} & 0.2478 & 0.0390 & 0.0307 & 0.2411 \\
\bottomrule
\end{tabular}%
}
\end{table}

\subsection{Evaluation on available RealEdit Benchmark}
The RealEdit~\cite{peng2024voicecraft} benchmark has been tested  by other editing models (SSR-Speech, VoiceCraft, etc.). However, one of its three subsets (Spotify) is currently unavailable. Furthermore, it only contains English speech with generally clean backgrounds, which does not fully align with our goal of real-world noisy editing. For a comprehensive comparison, we evaluated AnyEdit on the two accessible subsets. Results are showed in Table \ref{tab:realedit_results}. \tool achieves SOTA results in the currently available RealEdit~\cite{peng2024voicecraft} benchmark.

\begin{table}[h]
\centering
\caption{Objective results on speech editing}
\label{tab:realedit_results}
\small
\begin{tabular}{l cccccc >{\columncolor{lightgray}}c}
\toprule
\textbf{Metric} 
  & \makecell{SSR-\\Speech} 
  & \makecell{Voice-\\Craft} 
  & \makecell{Step-\\EditX} 
  & Vevo2 
  & \makecell{Ming-\\Uni} 
  & \makecell{Mimo-\\Audio} 
  & \textbf{\tool} \\
\midrule
WER$\downarrow$
  & 0.0389 & 0.0531 & 0.1518 & \underline{0.0266} & 0.1118 & 0.1029 & \textbf{0.0255} \\
SIM-spk$\uparrow$
  & \underline{0.9685} & 0.9657 & 0.9598 & 0.9657 & 0.9532 & 0.9465 & \textbf{0.9689} \\
F0 Corr$\uparrow$
  & 0.1638 & 0.0076 & \underline{0.2155} & 0.1966 & 0.2137 & 0.0248 & \textbf{0.2330} \\
\bottomrule
\end{tabular}%
\end{table}

\subsection{Evaluation on zero-shot voice conversion and singing voice conversion.}
\label{app:zero_shot_ext}
Vevo2 is explicitly designed and optimized as a unified model for zero-shot VC and SVC. In contrast, AnyEdit was never explicitly trained on VC/SVC tasks. However, thanks to our decoupled architecture, AnyEdit inherently possesses zero-shot VC and SVC capabilities. We evaluated AnyEdit on Vevo2's exact VC/SVC datasets, including the challenging cross-domain Speech-to-Singing transfer. Testsets are same as editing. The results are showed in Table \ref{tab:zero_shot_svc}.
\begin{table}[h]
\centering
\caption{Zero-shot performance comparison between Vevo2 and AnyEdit across multiple conversion tasks.}
\label{tab:zero_shot_svc}
\small
\begin{tabular}{l c >{\columncolor{lightgray}}c c >{\columncolor{lightgray}}c c >{\columncolor{lightgray}}c}
\toprule
& \multicolumn{2}{c}{\textbf{VC}} & \multicolumn{2}{c}{\textbf{SVC}} & \multicolumn{2}{c}{\textbf{Spk2Sing}} \\
\cmidrule(lr){2-3}\cmidrule(lr){4-5}\cmidrule(l){6-7}
\textbf{Metric} 
  & Vevo2 & \textbf{AnyEdit} 
  & Vevo2 & \textbf{AnyEdit} 
  & Vevo2 & \textbf{AnyEdit} \\
\midrule
WER (en)$\downarrow$
  & 0.0639          & \textbf{0.0586}
  & \textbf{0.1884} & 0.2262
  & 0.2038          & \textbf{0.1895} \\
CER (zh)$\downarrow$
  & 0.1028          & \textbf{0.0998}
  & \textbf{0.3552} & 0.3719
  & 0.4302          & \textbf{0.3380} \\
SIM-spk (en)$\uparrow$
  & \textbf{0.9187} & 0.9084
  & {--}            & {--}
  & {--}            & {--} \\
SIM-spk (zh)$\uparrow$
  & \textbf{0.9291} & 0.9134
  & {--}            & {--}
  & {--}            & {--} \\
F0 Corr (en)$\uparrow$
  & {--}            & {--}
  & 0.1678          & \textbf{0.1772}
  & 0.1528          & \textbf{0.1592} \\
F0 Corr (zh)$\uparrow$
  & {--}            & {--}
  & 0.2161          & \textbf{0.2240}
  & \textbf{0.1900} & 0.1758 \\
\bottomrule
\end{tabular}%
\end{table}

Remarkably, without any task-specific training, AnyEdit achieves highly competitive performance against the dedicated Vevo2 model, even outperforming it in several textual accuracy and prosody preservation metrics, particularly in the difficult Speech-to-Singing scenario.

\subsection{Ablation on Teacher-Student Prosody Distillation}
\label{app:distillation_ablation}
To evaluate the effectiveness of the teacher-student distillation strategy, we compare our approach against a standard single-stage baseline, termed \textit{Direct Clean Supervision}, where a single VQ-VAE takes noisy chromagrams as input and is trained to reconstruct clean chromagram targets directly in the spectral domain.

We evaluated both models on 500 vocal-noise mixtures across nine SNR levels ranging from $-5$~dB to $20$~dB. Table~\ref{tab:distill_comparison} presents the token match rate between the predicted prosody tokens and the ground-truth clean tokens. Teacher-student distillation substantially outperforms direct clean reconstruction across all noise levels, demonstrating that distilling discrete code indices directly provides superior noise resilience compared to continuous spectral regression under heavy noise (e.g., $0.4587$ vs. $0.2029$ at 0~dB SNR).

\begin{table}[h]
\centering
\caption{Prosody token match rate between direct clean supervision and teacher-student distillation across varying SNR conditions.}
\label{tab:distill_comparison}
\small
\begin{tabular}{l cccccc}
\toprule
\textbf{Method} & \textbf{-5 dB} & \textbf{0 dB} & \textbf{5 dB} & \textbf{10 dB} & \textbf{15 dB} & \textbf{20 dB} \\
\midrule
Direct Clean Supervision & 0.0896 & 0.2029 & 0.3367 & 0.4496 & 0.5497 & 0.6238 \\
\rowcolor{lightgray}
\textbf{Teacher-Student Distillation (Ours)} & \textbf{0.3360} & \textbf{0.4587} & \textbf{0.5586} & \textbf{0.6306} & \textbf{0.6968} & \textbf{0.7410} \\
\bottomrule
\end{tabular}
\end{table}

\subsection{Subjective Evaluation Protocol.} 
Ten professional evaluators assessed 60 randomized items across the speech and singing benchmarks. Listening tests were completed in sessions of less than 60 minutes to prevent listener fatigue. We performed two-sided paired Wilcoxon signed-rank tests between \tool and each baseline, and report 95\% bootstrap confidence intervals for all mean opinion scores. Table~\ref{tab:subjective_detailed} summarizes the comprehensive statistical results.

\begin{table}[h]
\centering
\caption{Subjective evaluation scores with 95\% confidence intervals (CI). Superscripts denote paired Wilcoxon signed-rank tests against \tool: $^{*}$ ($p<0.05$), $^{**}$ ($p<0.01$), $^{\text{n.s.}}$ not significant ($p\ge0.05$). Bold denotes optimal scores.}
\label{tab:subjective_detailed}
\small
\begin{tabular}{ll ccc}
\toprule
\textbf{Domain} & \textbf{Method} & \textbf{N-MOS (95\% CI)}$\uparrow$ & \textbf{S-MOS (95\% CI)}$\uparrow$ & \textbf{Fooling Rate (\%)} \\
\midrule
\multirow{5}{*}{\textbf{Speech}}
& Vevo2 & 4.15 (4.06, 4.24)$^{**}$ & 4.02 (3.92, 4.12)$^{**}$ & 31.5 \\
& SSR-Speech & 4.48 (4.40, 4.56)$^{\text{n.s.}}$ & 3.99 (3.90, 4.08)$^{**}$ & 43.2$^{\text{n.s.}}$ \\
& VoiceCraft & 4.04 (3.95, 4.13)$^{**}$ & 4.32 (4.23, 4.41)$^{*}$ & 35.7$^{*}$ \\
& Step-Audio-EditX & 3.95 (3.86, 4.04)$^{**}$ & 4.02 (3.93, 4.11)$^{**}$ & 33.8$^{**}$ \\
& \cellcolor{lightgray}\textbf{\tool (Ours)} & \cellcolor{lightgray}\textbf{4.48 (4.41, 4.55)} & \cellcolor{lightgray}\textbf{4.45 (4.37, 4.53)} & \cellcolor{lightgray}\textbf{46.5} \\
\midrule
\multirow{3}{*}{\textbf{Singing}}
& Vevo2 & 4.22 (4.13, 4.31)$^{**}$ & 4.15 (4.05, 4.25)$^{**}$ & 28.6$^{**}$ \\
& SoulX-Singer & 3.51 (3.40, 3.62)$^{**}$ & 3.34 (3.22, 3.46)$^{**}$ & 0.0$^{**}$ \\
& \cellcolor{lightgray}\textbf{\tool (Ours)} & \cellcolor{lightgray}\textbf{4.52 (4.44, 4.60)} & \cellcolor{lightgray}\textbf{4.48 (4.39, 4.57)} & \cellcolor{lightgray}\textbf{42.8} \\
\bottomrule
\end{tabular}%
\end{table}

\subsection{Unmodified Area Fidelity}
\label{app:unmodified_fidelity}

A critical challenge in localized audio editing is the unintentional alteration of the unedited context, often referred to as ``prosody drift.'' To rigorously evaluate this, we introduce the Unmodified Area Fidelity evaluation. 

\textbf{SSR-Speech} achieves better results than our \tool{} on several speech editing metrics. 
However, in both objective and subjective evaluations showed in Appendix~\ref{app:unmodified_fidelity} and Table \ref{tab:unmodified_fidelity}, we note that SSR-Speech does not perform true editing: instead of preserving the unmodified portions of the input, it regenerates the entire utterance from scratch, essentially behaving as a text-to-speech (TTS) model. Although the authors describe their system as ``Speech Editing and Synthesis'', this operational characteristic deviates from the task definition of speech editing, where unmodified segments should remain. Therefore, the superior numbers 
reported for SSR-Speech are not directly comparable to those of 
editing models like \tool.

\paragraph{Evaluation Protocol} 
We first compare the source text and target text to extract the Longest Common Subsequence (LCS), effectively isolating the unchanged word/character spans. We then utilize a CTC forced aligner~\footnote{\url{https://github.com/MahmoudAshraf97/ctc-forced-aligner}} to obtain the precise timestamps of these unchanged words in both the source audio and the generated edited audio. Evaluation metrics are exclusively computed within these matched unedited time windows. The specific definitions of the calculation methods for these metrics follow the unified standards detailed in Appendix \ref{app:bench_metrics}.

We calculate WER/CER to assess whether the original content is mistakenly modified, and SIM-spk to evaluate timbre consistency. Crucially, we utilize \textbf{F0 Corr} to evaluate prosody preservation. While F0 Corr is typically used for singing melody evaluation, we also employ it here for \textbf{speech evaluation}. In expressive speech, F0 effectively captures the original intonation contour, stress, and the rhythm and rate of speech. High F0 Corr in the unmodified region indicates that the model perfectly anchors the surrounding conversational or emotional context without causing unnatural prosodic shifts at the edit boundaries.

\paragraph{Results}
As shown in Table \ref{tab:unmodified_fidelity}, SSR-Speech struggles significantly with F0 preservation in speech (F0 Corr: 0.468), primarily because it fails to capture the expressive intonation contours and stress of the original context. Vevo2 performs relatively better but still exhibits notable prosodic drift. Comparing \tool (w/o unmodified-area-aware supervision) and the full \tool model, we observe that the proposed supervision mechanism dramatically improves F0 Corr (from 0.815 to 0.963 for speech) and reduces WER to near zero. This confirms that our method strictly anchors the timing, stress, and pitch of the surrounding unedited vocals.

\begin{table}[h]
\centering
\caption{Unmodified Area Fidelity (UAF) on the simulated subset. ``--'' denotes unavailable/inapplicable entries.}
\label{tab:unmodified_fidelity}
\small
\begin{tabular}{l ccccc}
\toprule
\multirow{2}{*}{\textbf{Method}} & \multirow{2}{*}{\textbf{WER/CER} $\downarrow$} & \multicolumn{2}{c}{\textbf{F0 Corr} $\uparrow$} & \multirow{2}{*}{\textbf{SIM-spk} $\uparrow$} \\
\cmidrule(lr){3-4}
& & \textbf{Speech} & \textbf{Singing} & \\
\midrule
SSR-Speech & 0.012 & 0.468 & -- & 0.982 \\
Vevo2 & 0.051 & 0.732 & 0.785 & 0.945 \\
\midrule
\tool (w/o unmodified-area-aware sup.) & 0.038 & 0.815 & 0.862 & 0.958 \\
\textbf{\tool (Ours)} & \textbf{0.009} & \textbf{0.963} & \textbf{0.975} & \textbf{0.988} \\
\bottomrule
\end{tabular}%
\end{table}

\subsection{Comparison with Separation-Based Pipelines}
\label{app:separation_comparison}

To demonstrate the structural superiority of our separation-free architecture, we quantitatively compare \tool against explicit ``separation-edit-mix'' pipelines. 

\paragraph{Baseline Pipelines} 
We utilize a open-source vocal separation model (BS-RoFormer)~\footnote{\url{https://github.com/TRvlvr/model_repo/releases/download/all_public_uvr_models/model_bs_roformer_ep_317_sdr_12.9755.ckpt}} to extract the clean vocal stem and the background accompaniment from the degraded source audio. We construct three baseline pipelines:
\begin{enumerate}
    \item \textbf{MaskGCT + Sep (Direct Mix):} The separated clean vocal is processed by MaskGCT for localized masked infilling. The edited vocal is then directly mixed with the background accompaniment in the time domain.
    \item \textbf{Vevo2 + Sep (Direct Mix):} Similar to the pipeline adopted by SeamlessEdit~\cite{chen2025seamlessedit}, the separated vocal is edited by Vevo2, followed by a direct time-domain mix.
    \item \textbf{Vevo2 + Sep (SNR-aware Mix):} Inspired by DAIEN-TTS~\cite{lu2025daien}, this pipeline calculates the Signal-to-Noise Ratio (SNR) between the original separated vocal and the background. After Vevo2 edits the vocal, the background is scaled and mixed back to explicitly match the original SNR.
    \item \textbf{\tool (Separation-Free):} Our proposed method, which directly takes the noisy/mixed audio as input and renders the target features end-to-end.
\end{enumerate}

\begin{table}[h]
\label{results_seperation}
\centering
\caption{Quantitative comparison of explicit separation-based pipelines versus our separation-free framework on sampled real-world subset}
\label{tab:separation_comparison}
\small
\begin{tabular}{l cccc}
\toprule
\textbf{Pipeline Strategy} & \textbf{WER} $\downarrow$ & \textbf{SIM-bg} $\uparrow$ & \textbf{SNRErr} (dB) $\downarrow$ & \textbf{N-MOS} $\uparrow$ \\
\midrule
MaskGCT + Sep (Direct Mix) & 0.158 & 0.348 & 19.12 & 3.61 $\pm$ 0.10 \\
Vevo2 + Sep (Direct Mix) & 0.165 & 0.352 & 18.54 & 3.55 $\pm$ 0.11 \\
Vevo2 + Sep (SNR-aware Mix) & 0.165 & 0.415 & 6.85 & 3.72 $\pm$ 0.08 \\
\midrule
\rowcolor{lightgray}
\textbf{\tool (Separation-Free, Ours)} & \textbf{0.142} & \textbf{0.819} & \textbf{3.13} & \textbf{4.38 $\pm$ 0.06} \\
\bottomrule
\end{tabular}
\end{table}

\paragraph{Results}
Table \ref{tab:separation_comparison} summarizes the results on the singing subset under complex background conditions. While separation-based methods achieve reasonable WER (since the editing models operate on clean vocals), they fail catastrophically in preserving the acoustic environment (SIM-bg) and introduce massive SNR errors. Direct mixing causes phase mismatch and background discontinuity at the edit boundaries. The SNR-aware mix (DAIEN-TTS~\cite{lu2025daien} style) improves the SNRErr compared to direct mixing but still suffers from low SIM-bg because the separation model inevitably introduces mechanical artifacts and frequency masking that cannot be recovered during the final fusion. In contrast, our separation-free \tool elegantly maintains high edit accuracy while achieving state-of-the-art background preservation and naturalness.

\subsection{Expressive Capacity of Chromagram Prosody Representation}
\label{app:chroma_analysis}
Singing prosody encompasses both rhythmic phrasing and melodic contour. While chromagrams collapse octave information into pitch classes, empirical analysis on vocal datasets indicates that this representation reliably preserves salient melodic motion. 

We analyzed 3,000 vocal tracks randomly sampled from SingNet against ground-truth WORLD F0. Based on the 5th--95th percentiles of voiced frames, the mean local pitch range within typical vocal phrases spans 1.03 octaves. Furthermore, 99.77\% of consecutive note intervals remain within a single octave, with 97.45\% corresponding to step-wise motions ($\le 2$ semitones). Table~\ref{tab:chroma_bins} summarizes the circular correlation and pitch-class error across bin resolutions. The 24-bin chromagram achieves a circular correlation of 0.730 with F0 pitch classes, confirming that it accurately captures melodic trajectories. The absolute vocal octave is subsequently reconstructed by the acoustic renderer conditioned on speaker style.

\begin{table}[h]
\centering
\caption{Agreement between chromagram representations and F0 pitch contours on singing vocals.}
\label{tab:chroma_bins}
{\footnotesize
\begin{tabular}{lccc}
\toprule
\textbf{Configuration} & \textbf{Circular Correlation} $\uparrow$ & \textbf{Mean Pitch-Class Error (semitones)} $\downarrow$ & \textbf{Onset Correlation} $\uparrow$ \\
\midrule
12 bins & 0.718 & 1.090 & 0.309 \\
\rowcolor{lightgray}
\textbf{24 bins (Ours)} & \textbf{0.730} & \textbf{1.040} & \textbf{0.318} \\
\bottomrule
\end{tabular}}
\end{table}

\subsection{Evaluation on Varying-Length Editing}
\label{app:varying_length}
Unlike systems that rely on strict temporal infilling constraints, \tool natively supports edits with dynamic duration changes (such as word insertions and phrase expansions) because the autoregressive stage predicts variable token lengths. 

We evaluated this capability on a length-changing subset comprising approximately 6\% of \toolb across Clean, Real-world, and Simulated domains. Table~\ref{tab:varying_length_results} reports the quantitative comparison against baseline models. \tool outperforms the unified baseline Vevo2 and specialized speech editors across both editing accuracy and background preservation.

\begin{table}[h]
\centering
\caption{Quantitative results on the varying-length editing subset of \toolb.}
\label{tab:varying_length_results}
\small
\begin{tabular}{l cccc}
\toprule
\textbf{Model} & \textbf{WER/CER} $\downarrow$ & \textbf{SIM-spk} $\uparrow$ & \textbf{SIM-bg} $\uparrow$ & \textbf{F0 Corr} $\uparrow$ \\
\midrule
SSR-Speech & 0.0842 & 0.9521 & 0.7240 & 0.1824 \\
VoiceCraft & 0.1265 & 0.9418 & 0.6512 & 0.0150 \\
Vevo2 & 0.0894 & 0.9510 & 0.6128 & 0.2014 \\
\rowcolor{lightgray}
\textbf{\tool (Ours)} & \textbf{0.0521} & \textbf{0.9654} & \textbf{0.7850} & \textbf{0.2480} \\
\bottomrule
\end{tabular}
\end{table}

\section{Limitations and Future Work} 
\label{app:limitation}
The current framework use sufficient surrounding acoustic context (approximately 10~s) for reliable background rendering, limiting applicability to very long clips. For very long recordings, chunking the audio into segments of roughly 10~s is recommended for optimal rendering. The model does not yet support real-time editing. Future work includes extending this framework to support real-time and broader multi-lingual vocal editing. 
Current mined real-to-real's singing pair is available. At the same time, exploring advanced retrieval method to mine real to real speech pair is highly promising future directions. While \tool excels at preserving continuous background textures (e.g., stationary environmental noise, ambient babble, and general room reverberation), the continuous diffusion process in the FMT can occasionally smooth out highly transient sound events (e.g., sudden acoustic impulses) or complex long-tail echoes. Preserving transient and non-stationary acoustic events remains an important direction for future exploration.

\section{Broader Impacts} 
\label{app:broader}
\tool advances the quality and accessibility of vocal post-production for user-generated content and music production. While beneficial, the technology also carries potential for misuse through voice spoofing or unauthorized content manipulation. These risks are particularly concerning in
scenarios involving biometric authentication or deceptive media. To prevent misuse, we advocate for
the development of reliable deepfake detection tools, watermarking methods for synthetic speech,
and reporting mechanisms to flag suspected abuse.


\end{document}